\documentclass[sigconf]{acmart}
\usepackage{booktabs}
\usepackage{pifont}
\usepackage{graphicx}
\usepackage{xcolor}
\usepackage{subcaption}
\usepackage{tikz}

\newcommand{\flowmark}[1]{%
  \tikz[baseline=(char.base)]{
    \node[shape=circle, fill=black, text=white, inner sep=0.6pt, font=\scriptsize\bfseries] (char) {#1};
  }%
}
\AtBeginDocument{%
  }

\setcopyright{none} 
\copyrightyear{2026}
\acmYear{2026}
\acmDOI{XXXXXXX.XXXXXXX}

\acmConference[MICRO 2026]{The 58th IEEE/ACM International Symposium on Microarchitecture}{October 31--November 04, 2026}{Athens, Greece}
\acmISBN{978-X-XXXX-XXXX-X/XX/XX}

\begin{document}

%%
%% The "title" command has an optional parameter,
%% allowing the author to define a "short title" to be used in page headers.
\title{C2P-Cache: Scalable GPU L1 Cache Sharing via Concurrent Candidate Pruning}
% \subtitle{\normalsize{MICRO 2026 Submission
%     \textbf{\#447} -- Confidential Draft -- Do NOT Distribute!!}}

\title{C2P-Cache: Scalable GPU L1 Cache Sharing via Concurrent Candidate Pruning}

\author{%
Hanqing Li,
Lizhou Wu\textsuperscript{*},
Tiejun Li\textsuperscript{*},
Sheng Ma\textsuperscript{*},
Hanzhi Xun\textsuperscript{*},
Jianmin Zhang,
Yuhan Tang,
Jixuan Tang,
and Xuchao Xie
}

\affiliation{%
  \institution{National University of Defense Technology}
  \city{Changsha}
  % \city{}
  % \country{}
  \country{China}
}

\email{%
{lihanqing23013, lizhou.wu, tjli, masheng, xunhanzhi,
jmzhang, tangyuhan, tangjixuan19, xiexuchao}@nudt.edu.cn
}

\renewcommand{\shortauthors}{Li et al.}
%% The "author" command and its associated commands are used to define
%% the authors and their affiliations.
%% Of note is the shared affiliation of the first two authors, and the
%% "authornote" and "authornotemark" commands
%% used to denote shared contribution to the research.
%\author{\normalsize{MICRO 2026 Submission
 %   \textbf{\#NaN} -- Confidential Draft -- Do NOT Distribute!!}}

%%
%% By default, the full list of authors will be used in the page
%% headers. Often, this list is too long, and will overlap
%% other information printed in the page headers. This command allows
%% the author to define a more concise list
%% of authors' names for this purpose.

%%
%% The abstract is a short summary of the work to be presented in the
%% article.

%%%%%% -- PAPER CONTENT STARTS-- %%%%%%%%

\begin{abstract}

Modern GPUs rely on private per-SM L1 caches and a shared L2 cache, but this organization obscures cross-SM reuse: an L1 miss is typically forwarded to L2 even when the requested line already resides in a peer L1 cache, leading to redundant L2 access. Prior GPU L1-sharing designs attempt to recover such reuse through exact or broad remote-hit searches, which become increasingly difficult to scale and can interfere with the critical L1 miss path under high concurrency. %miss handling as more caches participate and more misses arrive concurrently.
We observe that eliminating redundant L2 accesses does not require exact, chip-wide knowledge of private L1 contents. 
Instead, it requires only sufficient visibility to sharply narrow down a small set of candidate caches, leaving exact confirmation to a much smaller number of L1s. 

Based on this insight, we propose C2P-Cache, a scalable GPU L1-sharing mechanism that transforms remote-hit discovery from a chip-wide exact search problem into a lightweight filtering-and-confirmation process.
C2P-Cache maintains compact Bloom-filter-based snapshots of private L1 tags, performs parallel chip-wide candidate filtering, and selectively probes only a small number of likely peer caches. To sustain high concurrency, C2P-Cache organizes filtering as bit-sliced matching over a banked and replicated snapshot matrix, enabling efficient, parallel processing of many concurrent misses without interfering with normal L1 accesses.
Across a wide range of GPU workloads, C2P-Cache improves instructions per cycle (IPC) by up to 49.7\% and by 23.5\% on average for applications with high remote-L1 reuse and strong sensitivity to L2 latency, demonstrating that lightweight, scalable filtering can effectively unlock cross-SM reuse with modest overhead.
%.maintains compact snapshots of private L1 tags, performs approximate chip-wide filtering on each L1 miss, and selectively probes only the remaining candidate peer L1 caches for exact confirmation. C2P-Cache further supports parallel remote-hit identification with a banked and replicated Snapshot Matrix. Across diverse GPU workloads, C2P-Cache improves IPC by up to 49.7\%, with an average improvement of 23.5\% for applications that exhibit abundant remote-L1 reuse and high sensitivity to the L2 access path.

\end{abstract}
%%
%% The code below is generated by the tool at http://dl.acm.org/ccs.cfm.
%% Please copy and paste the code instead of the example below.
%%
%\begin{CCSXML}
%<ccs2012>
% <concept>
%  <concept_id>00000000.0000000.0000000</concept_id>
%  <concept_desc>Do Not Use This Code, Generate the Correct Terms for Your Paper</concept_desc>
%  <concept_significance>500</concept_significance>
% </concept>
% <concept>
%  %<concept_id>00000000.00000000.00000000</concept_id>
%  <concept_desc>Do Not Use This Code, Generate the Correct Terms for Your Paper</concept_desc>
%  <concept_significance>300</concept_significance>
% </concept>
% <concept>
%  %<concept_id>00000000.00000000.00000000</concept_id>
%  <concept_desc>Do Not Use This Code, Generate the Correct Terms for Your Paper</concept_desc>
%  <concept_significance>100</concept_significance>
% </concept>
% <concept>
 % <concept_id>00000000.00000000.00000000</concept_id>
%  <concept_desc>Do Not Use This Code, Generate the Correct Terms for Your Paper</concept_desc>
%  <concept_significance>100</concept_significance>
% </concept>
%</ccs2012>
%\end{CCSXML}

%\ccsdesc[500]{Do Not Use This Code~Generate the Correct Terms for Your Paper}
%\ccsdesc[300]{Do Not Use This Code~Generate the Correct Terms for Your Paper}
%\ccsdesc{Do Not Use This Code~Generate the Correct Terms for Your Paper}
%\ccsdesc[100]{Do Not Use This Code~Generate the Correct Terms for Your Paper}

%%
%% Keywords. The author(s) should pick words that accurately describe
%% the work being presented. Separate the keywords with commas.
\keywords{GPU architecture, Bloom filter, cache sharing}

\maketitle

\begingroup
\renewcommand{\thefootnote}{*}
\footnotetext{Corresponding authors.}
\endgroup

%%%%%%%%%%%%%%%%%%%%%%%%%%%%%%%%%%%%%%%%%%%%%%%%%%%%%%%%%%%%%%%%%%%
\section{Introduction}

Modern GPUs rely on massive thread-level parallelism to deliver high throughput, yet memory behavior remains a key performance limiter for many applications \cite{GPUMEMORYWALL}. A typical GPU employs private L1 caches for streaming multiprocessors (SMs) and a shared L2 cache as the last on-chip cache before global memory \cite{A100_white_paper,H100_white_paper}. While this organization provides fast local access, it also hides reuse opportunities across SMs: a cache line already resident in a SM's private L1 cache remains invisible to others \cite{l1cacheredundancy1,l1cacheredundancy2}. 
As a result, an L1 miss is conventionally forwarded to the shared L2 cache even when the requested data already exists in a peer private L1 cache, leading to redundant L2 access. These unnecessary accesses consume L2 cache bandwidth, increase on-chip traffic, and prolong miss latency by unnecessarily forcing requests onto the L2 path.
Our characterization shows that redundant L2 accesses arise across a broad range of GPU workloads rather than being confined to a few corner cases. More importantly, their performance impact depends jointly on two factors: the availability of remote-L1 reuse and the application's sensitivity to L2 latency. 
%how often an L1 miss could be satisfied by a peer private L1 cache, and how strongly the application is affected by the L2 access path. 
When both are high, eliminating redundant L2 accesses becomes a critical opportunity to improve overall memory-system efficiency.
%Cross-SM cache sharing is therefore most beneficial when abundant remote-L1 reuse opportunities coincide with high sensitivity to L2 latency. This observation suggests that eliminating redundant L2 accesses is not merely a corner-case optimization, but a meaningful opportunity to improve GPU memory-system efficiency. It also implies that any practical sharing mechanism must keep its own overhead low, since the benefit varies substantially across workloads.

Enabling effective cross-SM L1 cache sharing, however, is fundamentally challenging. A practical solution must satisfy three key requirements. First, it must provide sufficiently broad visibility to capture reuse opportunities across all SMs, rather than being confined to a small local domain. Second, it must introduce minimal interference to the critical L1 miss path, avoiding additional pressure on latency-sensitive cache accesses. Third, it must sustain high concurrency under many simultaneous misses, without turning remote-hit identification itself into a throughput bottleneck.
These requirements highlight a central challenge: how to efficiently identify useful peer-L1 candidates at chip scale, without relying on expensive global search or introducing significant overhead.

Prior work has explored GPU L1 cache sharing using techniques such as predictor-guided broadcast \cite{sharingl1-CCD,sharingl1-MeshL1sharing}, centralized metadata lookup \cite{sharingl1-ATA,sharingl1-L1.5Dcache}, and traversal-based remote probing \cite{sharingl1-RING}. While these approaches improve reuse to some extent, they typically rely on relatively broad or exact searches to determine whether a peer L1 cache can satisfy a miss.
Such designs suffer from fundamental limitations. Broadcast-based approaches incur excessive probing overhead and depend heavily on prediction accuracy. Centralized metadata structures constrain lookup throughput and scalability. Traversal-based probing limits discovery latency and concurrency due to serialized exploration. More generally, as the number of SMs and concurrent misses increases, these approaches struggle to scale, often turning remote-hit discovery itself into a performance bottleneck.
These limitations stem from a common root cause: they treat remote-L1 lookup as a chip-wide exact-discovery problem, requiring either broad probing or centralized coordination.

In this work, we observe that eliminating redundant L2 accesses does not require exact chip-wide knowledge of all private L1 contents. Instead, it requires only sufficient visibility to quickly narrow down a small set of likely candidate caches, leaving exact confirmation to a much smaller subset. This insight transforms remote-L1 handling from an exact-discovery problem into a lightweight filtering-and-confirmation problem.
Guided by this observation, we propose \textbf{C2P-Cache}, a scalable GPU L1-sharing mechanism that uses compact Bloom-filter-based snapshots to provide approximate, chip-wide visibility of private L1 contents. Upon an L1 miss, C2P-Cache performs parallel filtering to prune most impossible remote locations and selectively probes only a small number of candidate peer L1 caches for exact confirmation, while preserving the baseline L2 access path on misses.
%maintains compact Bloom-filter snapshots of private L1 tags and consults them before an L1 miss is sent to L2, quickly ruling out remote caches that cannot contain the requested cache line. Only the small set of remaining candidate peer L1 caches is then selectively probed for exact confirmation. If none of them supplies the block, the request simply follows the unchanged baseline L2 path; otherwise, a remote hit is returned through the normal refill path. In this way, C2P-Cache confines exact remote-hit confirmation to a small candidate set selected by Bloom-filter-guided filtering.
To support high concurrency, C2P-Cache organizes filtering as bit-sliced matching over a banked and replicated Snapshot Matrix, enabling parallel processing of many concurrent misses without interfering with normal L1 accesses. By decoupling candidate generation from exact probing and avoiding global exact search, C2P-Cache achieves scalable cross-SM cache sharing with broad visibility, low overhead, and high concurrency. The main contributions of this paper are as follows:

%C2P-Cache is designed around the three requirements above. First, to provide broad reuse visibility, it maintains Bloom-filter snapshots for all private L1 caches, enabling candidate generation from a chip-wide perspective rather than restricting sharing to a small local domain. Second, to keep overhead and cache interference low, it uses compact Bloom-filter queries to filter out most impossible remote locations before any exact probe, so that only a small candidate set needs confirmation while unresolved misses continue along the baseline L2 path. Third, to sustain high concurrency, it organizes Bloom-filter matching as bit-sliced operations over a banked and replicated Snapshot Matrix, allowing misses from across the chip to be processed in parallel and making remote-hit identification less likely to become a throughput bottleneck.

\begin{itemize}
    \item We propose \textbf{C2P-Cache}, a GPU L1 cache sharing mechanism that replaces expensive chip-wide exact remote-hit discovery with low-cost approximate filtering followed by selective exact confirmation.
    \item We design a \textbf{snapshot-based remote-hit identification framework} that maintains a compact snapshot of private L1 tags, filters remote-hit candidates before exact probing, and preserves the baseline L2 access path on unsuccessful remote lookups.
    \item We develop a \textbf{high-concurrency matching architecture} that enables parallel remote-hit identification across SMs via Bloom-filter matching over a banked and replicated Snapshot Matrix.
    \item We conduct a comprehensive evaluation on diverse GPU workloads and show that C2P-Cache improves instructions per cycle (IPC) by up to 49.7\%, with an average improvement of 23.5\% for applications that exhibit abundant remote-L1 reuse and high sensitivity to L2 latency.
\end{itemize}

\section{Background}
\subsection{GPU Architecture}
\begin{figure}[t]
    \centering
    \includegraphics[width=0.56\linewidth]{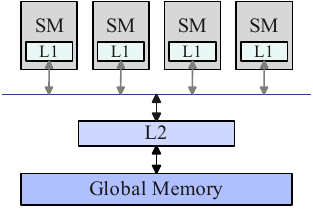}
    \caption{Simplified GPU memory hierarchy.}
    \label{fig:gpu-memory-hierarchy}
\end{figure}

Modern GPUs employ a hierarchical memory system with different capacity and latency trade-offs. Figure~\ref{fig:gpu-memory-hierarchy} illustrates the simplified hierarchy considered in this work. A GPU contains multiple SMs, each with a private L1 cache. Beyond the SM-private L1, all SMs share a unified L2, the last on-chip cache before off-chip global memory. Upper levels are smaller and faster, whereas lower levels provide larger capacity at higher access cost \cite{V100_white_paper,A100_white_paper,H100_white_paper,GB200_white_paper,cuda_programming_guide}. For memory-intensive GPU workloads, performance is often affected by how many requests reach global memory \cite{GPUMEMORYWALL,modern_gpu_modeling,medic_gpu_memory}. Once requests miss in the on-chip cache hierarchy and are forwarded to global memory, the access cost increases significantly and can become a major bottleneck. Therefore, reducing unnecessary accesses to the lower end of the memory hierarchy is critical to improving overall performance.

\subsection{Bloom Filter}

\begin{figure}[th]
    \centering
    \includegraphics[width=0.8\linewidth]{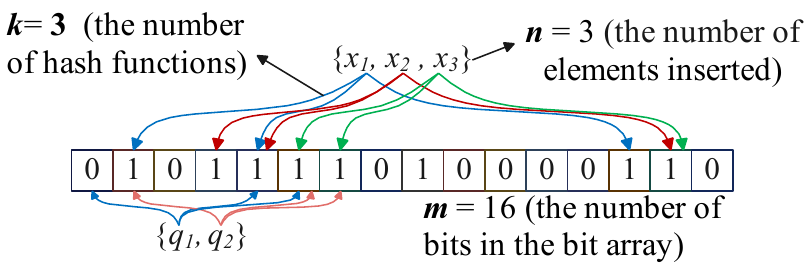}
    \caption{An example of Bloom-filter insertion and query.}
    \label{fig:bloom-filter}
\end{figure}

% We use BF as a low-cost pruning tool.
A Bloom filter (BF) \cite{bloomfilter} is a space-efficient probabilistic data structure for approximate membership testing. It is commonly used as a lightweight filter to determine whether an element may belong to a set before performing an exact lookup. In general, a Bloom filter consists of an $m$-bit array and $k$ hash functions, where $m$ denotes the filter size and $k$ denotes the number of hash functions. After inserting $n$ elements into the filter, membership queries can be answered approximately by checking the bit positions selected by these hash functions. Figure~\ref{fig:bloom-filter} shows an illustrative example with $m=16$, $k=3$, and $n=3$.

To insert an element, the Bloom filter applies the hash functions to the element and sets the corresponding bit positions in the array to 1. To query an element, it checks for the same bit positions. As illustrated in Figure~\ref{fig:bloom-filter}, inserted elements such as $x_1$, $x_2$, and $x_3$ set multiple bits in the array. For a query element, if any checked bit is 0, the query returns a \emph{miss}; for example, $q_1$ is a \emph{miss} because one of its mapped bits is 0. Otherwise, the query returns a \emph{hit}; in the figure, $q_2$ is a \emph{hit} because all of its mapped bits are 1. As a probabilistic structure, a Bloom filter may produce false positives, but it does not produce false negatives.

This asymmetry makes Bloom filters useful for filtering unnecessary lookups before costly exact accesses. A \emph{miss} can safely terminate the lookup early, whereas a \emph{hit} only indicates possible membership and still requires verification. Thus, Bloom filters provide an efficient way to prune unnecessary accesses.

\section{Motivation}
\label{sec:motivation}
\subsection{Redundant L2 Accesses}

In conventional GPUs, each SM has a private L1 cache whose contents are invisible to other SMs. Consequently, an L1 miss is always sent to the shared L2 cache, even when the requested data already resides in another SM's L1 cache. We call such cases \emph{redundant L2 accesses}, as the L2 access would be unnecessary if remote-L1 reuse could be exploited.

\begin{figure}[t]
    \centering
    \includegraphics[width=0.8\linewidth]{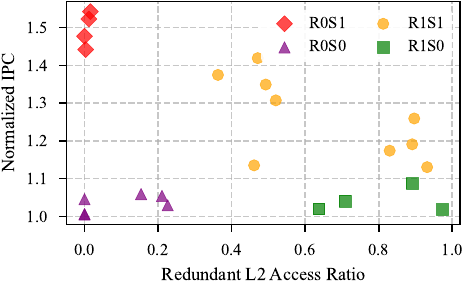}
    \caption{Redundant L2 accesses across GPU applications, grouped by redundant-access ratio and L2-latency sensitivity.}
    \label{fig:motivation_redundant_l2}
\end{figure}

Previous work has revealed such opportunities from different perspectives \cite{l1cacheredundancy1,l1cacheredundancy2,sharingl1-RING,sharingl1-MeshL1sharing,sharingl1-colab}. Here, we quantify their prevalence across various GPU applications. Figure~\ref{fig:motivation_redundant_l2} characterizes each application by the redundant L2 access ratio and the normalized IPC improvement when L2 latency is reduced from 200 to 50 cycles with L1 latency fixed at 20 cycles, indicating sensitivity to L2 latency. Using thresholds of 30\% redundancy and 1.1 sensitivity, we classify applications into four groups: R1S1, R1S0, R0S1, and R0S0. %These thresholds are chosen to distinguish workloads with clearly significant redundancy and sensitivity trends. 
Among them, R1S1 applications exhibit abundant remote-L1 reuse opportunities and high sensitivity to L2 latency.

The thresholds are characterization-only natural break points in Figure~\ref{fig:motivation_redundant_l2}, not runtime policy parameters. They identify workloads with substantial peer-L1 reuse and clear L2-latency sensitivity, two traits widely studied in GPU cache-management research~\cite{scheduling4,sharingl1-L1.5Dcache,medic_gpu_memory}. R1S1 cases include graph and tree traversal, dense linear algebra, stencils, convolutions, and producer-consumer or tile-overlap reuse; Discussion further discusses their workload relevance.

Figure~\ref{fig:motivation_redundant_l2} reveals two key insights. First, GPU applications exhibit redundant L2 accesses to widely varying degrees. Second, redundancy alone is not strongly coupled with performance, and reducing L2 latency benefits only a subset of workloads. Therefore, cache sharing presents the clearest optimization opportunity for R1S1 applications, while its negative impact on the other groups should be carefully minimized.

\noindent\fbox{%
\parbox{0.98\linewidth}{%
\textbf{Insight \#1:} In principle, GPU L1 cache sharing can benefit many applications, but its overhead must be carefully minimized to avoid harming those that benefit little from sharing.
}%
}

\subsection{Limitations of Existing L1 Cache Sharing}

\begin{figure}[t]
    \centering

    \begin{subfigure}[t]{0.48\linewidth}
        \centering
        \includegraphics[width=\linewidth]{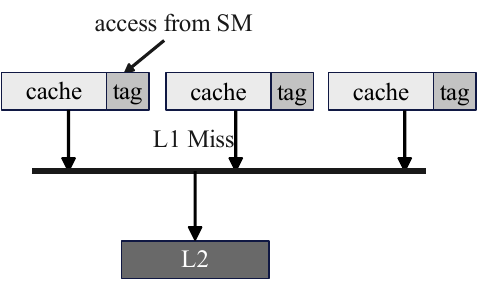}
        \caption{Traditional cache access}
        \label{fig:motivation_traditional}
    \end{subfigure}
    % \hfill
    \begin{subfigure}[t]{0.48\linewidth}
        \centering
        \includegraphics[width=\linewidth]{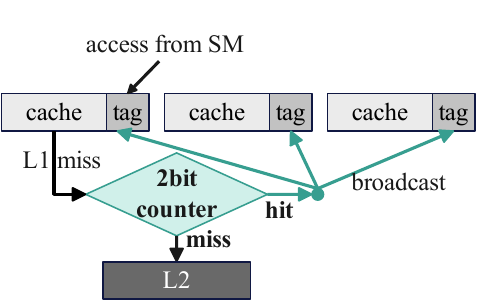}
        \caption{CCD \cite{sharingl1-CCD} cache sharing}
        \label{fig:motivation_ccd}
    \end{subfigure}

    % \vspace{2pt}

    \begin{subfigure}[t]{0.48\linewidth}
        \centering
        \includegraphics[width=\linewidth]{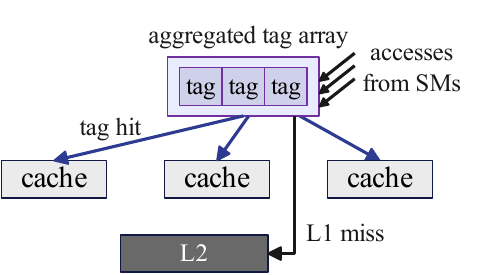}
        \caption{ATA \cite{sharingl1-ATA} cache sharing}
        \label{fig:motivation_ata}
    \end{subfigure}
    % \hfill
    \begin{subfigure}[t]{0.48\linewidth}
        \centering
        \includegraphics[width=\linewidth]{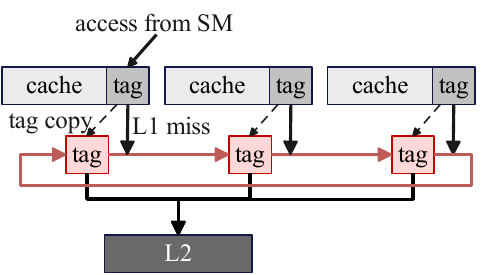}
        \caption{RING \cite{sharingl1-RING} cache sharing}
        \label{fig:motivation_ring}
    \end{subfigure}

    \caption{Comparison of L1 cache sharing designs.}
    \label{fig:motivation_existing_sharing}
\end{figure}

Figure~\ref{fig:motivation_existing_sharing} compares the conventional private-L1 design with three representative L1 cache sharing schemes. In the conventional design shown in Fig.~\ref{fig:motivation_existing_sharing}(a), each SM can access only its local L1 cache. As a result, once an L1 miss occurs, the request must proceed to L2 even if the requested data already resides in another SM's L1 cache. We select three effective prior L1 sharing schemes to illustrate the trade-offs that limit existing designs.

As shown in Fig.~\ref{fig:motivation_existing_sharing}(b), CCD~\cite{sharingl1-CCD} uses a 2-bit saturating counter per cluster to predict whether an L1 miss is likely to be redundant, similar to a branch predictor~\cite{2bitpred}. If so, it broadcasts the request to peer L1 caches in the cluster. This design has three main limitations. First, it exploits only intra-cluster reuse, so its sharing scope shrinks as SM count grows. Second, broadcast lookup adds substantial pressure on peer L1 caches. Third, CCD depends on predictor accuracy; ineffective predictions trigger many unnecessary L1 lookups.

Figure~\ref{fig:motivation_existing_sharing}(c) shows ATA~\cite{sharingl1-ATA}, which decouples tags from private L1 caches within a cluster and aggregates them into a separate array. A cache request first accesses this aggregated tag array, whose comparison result determines whether the request goes to the local L1, a remote L1, or L2. Although this design avoids explicit probing, it funnels all sharing requests through the aggregated tag array. As a result, its lookup throughput and parallelism are constrained by that structure. ATA is also limited to sharing within a cluster.

Figure~\ref{fig:motivation_existing_sharing}(d) shows a ring-based design~\cite{sharingl1-RING}, where L1 miss requests traverse an interconnect ring to probe remote caches. To reduce interference with normal cache accesses, RING replicates cache tags as shadow tags and accesses a remote cache only on a shadow-tag match. This avoids direct L1-cache interference and supports chip-wide sharing, but adds tag-storage overhead. Each miss also examines remote caches incrementally along the ring, increasing discovery latency and limiting concurrency. Under many L1 misses, requests and responses queue on the ring, making it a scalability bottleneck.

Overall, although CCD, ATA, and RING use different organizations, each sacrifices at least one property needed for scalable chip-wide sharing. CCD and ATA restrict sharing to a cluster, limiting sharing coverage as GPU scale increases, while ATA and RING limit lookup throughput or concurrency through centralized lookup and incremental ring traversal, respectively. Prior designs therefore trade off sharing scope, interference with normal L1 accesses, and lookup concurrency, making it difficult to simultaneously achieve broad coverage, low interference, and high scalability.

\noindent\fbox{%
\parbox{0.98\linewidth}{%
\textbf{Insight \#2:} 
Existing L1 cache sharing schemes improve reuse by trading off three key limitations: global visibility, interference with normal L1 accesses, and lookup concurrency. As a result, they scale poorly to large SM counts.
}%
}

% Although these schemes differ in organization, their limitations can be understood along three dimensions. First, the lack of global visibility, since their sharing support cannot efficiently cover all L1 caches at chip scale and is therefore restricted to a cluster. Second, the interfere with normal L1 accesses, because additional probing or metadata lookup introduces extra pressure beyond the normal local-cache path. Third, the limition of lookup concurrency, since the number of misses that can be handled in parallel is fundamentally bounded by the lookup structure (i.e the cache bank). These limitations do not necessarily appear together in every design, but prior sharing mechanisms typically trade off among them.

% \noindent\fbox{%
% \parbox{0.98\linewidth}{%
% \textbf{Key Observation:} Overall, existing L1 cache sharing schemes are constrained by three fundamental limitations: the lack of global visibility, interference with normal L1 accesses, and limited lookup concurrency.
% }%
% }

\section{C2P-Cache}
\subsection{C2P-Cache Overview}
\label{sec:design-overview}

\begin{figure}[t]
    \centering
    \includegraphics[width=\linewidth]{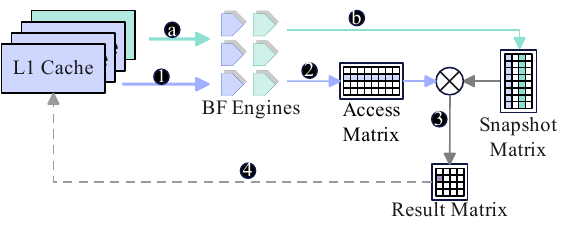}
    \caption{Overview of C2P-Cache.}
\label{fig:design_overview}
\end{figure}

C2P-Cache reduces redundant L2 accesses by checking if an L1 miss can be served by a peer L1 cache before forwarding it to L2. To avoid exact chip-wide remote-tag lookups, it maintains a lightweight BF-based snapshot, called the Snapshot Matrix, for each private L1 cache. These snapshots prune candidate L1 caches, while requests without remote hits fall back to the baseline L2 path. Figure~\ref{fig:design_overview} shows the overview of C2P-Cache. The design has two paths: a background update path that refreshes the Snapshot Matrix and a miss path that uses the Access Matrix, the Snapshot Matrix, and the Result Matrix to handle L1 misses.

On the update path, C2P-Cache refreshes the Snapshot Matrix in the background. \flowmark{a} C2P-Cache selects an L1 for update and sends its valid tags to the BF Engines. \flowmark{b} The generated results are written back to refresh the corresponding Snapshot Matrix column. By repeating this process across L1 caches, C2P-Cache keeps the Snapshot Matrix up to date without tracking every fill and eviction event.

On the miss path, \flowmark{1} the L1-miss tag is sent to the BF Engines and encoded into a BF row. The BF Engine is a dedicated hardware unit for generating the bit positions required by BF queries; in our design, it contains two BF hash units and one tag-mask unit. \flowmark{2} These rows are aggregated into the Access Matrix, which logically records miss queries. \flowmark{3} C2P-Cache then performs a Boolean Matrix multiplication between the Access Matrix and Snapshot Matrix to produce the Result Matrix. \flowmark{4} Finally, C2P-Cache prunes and probes the candidate L1 caches indicated by the Result Matrix. If a probe returns the requested cache line, C2P-Cache sends the line back through the on-chip interconnect and fills it into the requester's L1. Otherwise, the miss follows the baseline L2 path.

The BF Engines are lightweight index-generation lanes placed with the C2P miss-side controller next to the banked Snapshot Matrix, not inside the normal L1 hit datapath or storage arrays. The default design shares 128 BF/tag-mask engines between miss queries and background refresh. Miss queries have priority, while refresh work uses remaining engine bandwidth; Section~\ref{sec:eval-bf-lat-par} evaluates this parallelism.

\noindent\fbox{%
\parbox{0.98\linewidth}{%
\textbf{Insight \#3:}
The key idea of C2P-Cache is to exploit BF-based snapshots for lightweight, chip-wide candidate pruning, while using bit-level matching for high-concurrency candidate generation. By decoupling this pruning path from normal L1 accesses, C2P-Cache reduces interference and enables chip-wide L1 sharing with broad visibility, low interference, and high lookup concurrency, thereby providing better scalability.
}%
}

\subsection{Logical Matrix Formulation}
\label{sec:design-data-structures}

C2P-Cache is logically organized around three matrices: the Access Matrix, the Snapshot Matrix, and the Result Matrix. Together, they encode L1 misses and L1 tag states as a Boolean multiplication problem. Figure~\ref{fig:design_matrix_logic} illustrates the logical organization of the matrices.

\subsubsection{Access Matrix}
\label{sec:design-access-matrix}

The Access Matrix records the BF encodings of L1 miss requests. Logically, it has a dimension of $M \times m$, where $M$ is the number of missed requests and $m$ is the number of BF bit positions. Each row corresponds to one L1 miss and is extremely sparse: only a small number of queried positions are set by the encoding of the missed tag, while all remaining positions are zero. 

% Because the non-zero positions are generated directly by the BF Engines, C2P-Cache does not materialize each row as a full $m$-bit vector. Instead, each logical row is represented compactly by the requester SM identifier and the $k$ hash indices.

\subsubsection{Snapshot Matrix}
\label{sec:design-snapshot-matrix}

The Snapshot Matrix records BF-based tag snapshots for all L1 caches. Logically, it is an \(m \times N\) matrix, where \(N\) is the number of SMs, each column corresponds to an L1 cache, and each row corresponds to a BF bit position. More specifically, let $C_j$ denote the set of valid cache line tags currently stored in the L1 cache of SM $j$. The $j$-th column of the Snapshot Matrix is defined as the bitwise OR of the BF vectors of all tags in $C_j$:
\[
S_j = \bigvee_{t \in C_j} BF(t),
\]
where $BF(t)$ is the BF Engine encoding of the tag $t$. Compared with the Access Matrix, the Snapshot Matrix is much denser because each column aggregates the BF Engine encodings of all valid cache lines in one L1 cache.

\subsubsection{Result Matrix}
\label{sec:design-result-matrix}

\begin{figure}[t]
    \centering
    \includegraphics[width=1.0\linewidth]{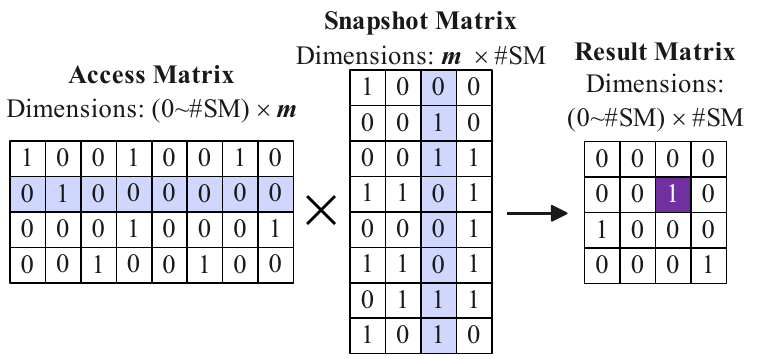}
    \caption{Logical view of C2P-Cache matching among the access matrix, snapshot matrix, and result matrix.}
    \label{fig:design_matrix_logic}
\end{figure}

The Result Matrix records the multiplication results between the Access Matrix and Snapshot Matrix. Logically, it has a dimension of $M \times N$, where each row corresponds to one L1 miss request, and each column corresponds to one L1 cache. Each entry is Boolean rather than numeric. A \textit{true} entry indicates that the corresponding L1 cache may contain the missed cache line, while a \textit{false} entry indicates that the cache definitely does not contain it. Thus, each row identifies the possible remote-L1 sources for an L1 miss. As illustrated in Figure~\ref{fig:design_matrix_logic}, this logical operation replaces numeric accumulation with bitwise intersection over the queried BF positions. After pruning the requester's own L1 cache and caches marked \textit{false}, C2P-Cache probes the remaining candidate caches.

\subsection{Access Matrix Encoding and Mapping}
\label{sec:design-matching}

Logically, C2P-Cache derives remote-L1 locations through Boolean matching between the Access Matrix and Snapshot Matrix. Rather than realizing this operation as a generic matrix-multiplication engine, C2P-Cache exploits the sparsity of miss-side queries and encodes each logical Access Matrix row as a set of queried positions. These positions determine which Snapshot Matrix rows must be accessed. This subsection focuses on how C2P-Cache designs the Access Matrix encoding to generate these lookup indices efficiently.

% This section describes how C2P-Cache represents each L1 miss on the miss path and uses that representation to locate possible peer L1 caches from the Snapshot Matrix. Rather than materializing a full logical Access Matrix row, C2P-Cache encodes each miss as a compact set of queried positions. These queried positions are then mapped to the corresponding Snapshot Matrix rows, whose matching results indicate which private L1 caches may contain the requested cache line.

% At a logical level, Result Matrix resembles a highly sparse-query--against--dense-snapshot operation. C2P-Cache exploits this structure by encoding each L1 miss (Access Matrix row) as a compact set of queried positions and organizing the Snapshot Matrix as row-indexed bitmaps over all SMs. As a result, candidate generation reduces to a few indexed row reads followed by a bitwise reduction.

\subsubsection{Compact Access Matrix Row Encoding}
\label{sec:design-query-encoding}

In the logical view, each L1 miss corresponds to a $5120$-bit access row consisting of a $1024$-bit tag-mask region and a $4096$-bit BF-hash region. Rather than relying on a pure BF encoding, C2P-Cache adopts a hybrid query encoding that combines one tag-mask function with multiple BF-hash functions on this logical row. Under the default configuration, each miss activates four queried positions in total: one in the tag-mask region and three in the BF-hash region.
The tag-mask component is motivated by three considerations. First, it naturally matches the lower 10 bits of the cache line tag and can be generated at very low hardware cost. Second, it improves discrimination under spatial locality, because nearby cache lines that share similar high tag bits can still be effectively distinguished by their lower tag bits, allowing the tag-mask term to serve as a fine-grained identifier. Third, it acts as a lightweight fingerprint that complements the BF-hash region. Prior work on fingerprint-enhanced BF variants has shown that compact fingerprint information can improve filtering quality and reduce false positives~\cite{CuckooFilter,bloom_finger_Sal}.

\begin{figure}[t]
    \centering
    \includegraphics[width=\linewidth]{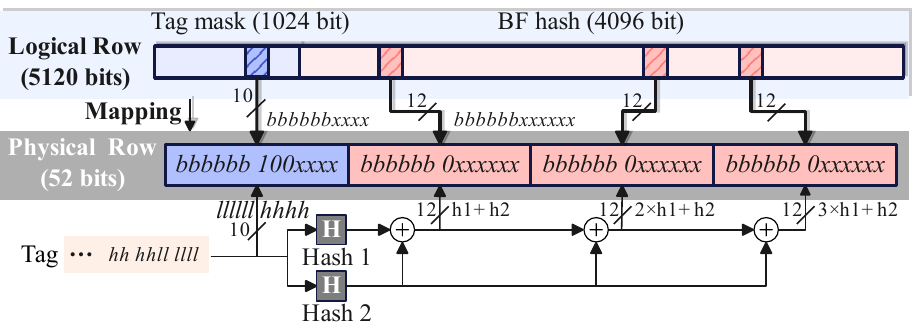}
    \caption{Logical-to-physical mapping of access matrix rows. }
    \label{fig:access_mapping}
\end{figure}

C2P-Cache uses a unified 13-bit format for all queried positions. A tag-mask position is encoded as $6$ bits + \texttt{100} + $4$ bits, whereas a BF-hash position is encoded as $6$ bits + \texttt{0} + $6$ bits. To generate the tag-mask position, C2P-Cache takes the lower 10 bits of the cache line tag, reverses their order, splits them into the upper six and lower four bits, and inserts the fixed pattern \texttt{100} between the two parts. To generate the BF-hash positions, C2P-Cache applies two base hash functions to the full tag, denoted as Hash 1 ($h_1$) and Hash 2 ($h_2$) in Figure~\ref{fig:access_mapping}, and derives three indices as
\[
b_1 = h_1 + h_2,\qquad
b_2 = 2h_1 + h_2,\qquad
b_3 = 3h_1 + h_2.
\]
Each derived BF-hash index is then split into the upper six and lower six bits, with a \texttt{0} inserted between them. This unified 13-bit format simplifies hardware implementation because both tag-mask and BF-hash positions can be decoded through the same addressing path when accessing the Snapshot Matrix.

Because the BF Engines directly produce these non-zero positions, C2P-Cache never materializes the logical row in hardware. It stores each miss as the physical row. In our implementation, this row is only $52$-bit wide, far smaller than a fully materialized $5120$-bit logical row. This compact representation avoids the storage overhead and format-conversion cost of materializing a sparse Access Matrix in hardware.
% \the\linewidth
% \begin{figure*}[t]
%     \centering
%     \includegraphics[width=\linewidth]{fig/design/AccessMatrixMapping.pdf}
%     \caption{Logical-to-physical mapping of Access Matrix rows. }
%     \label{fig:access_mapping}
% \end{figure*}

\subsubsection{Row Lookup and Bitwise Matching}
\label{sec:design-row-lookup}

Once an L1 miss has been encoded as a compact query entry containing four $13$-bit queried indices, C2P-Cache uses these indices to access the corresponding rows of the Snapshot Matrix. Logically, the Snapshot Matrix is organized as a row-indexed bitmap table with $m$ rows and $N$ bits per row, where each row corresponds to one queried position and the associated $N$-bit bitmap indicates which private L1 caches have that position set.

Given a miss query $Q_p=\{h_{p,1}, h_{p,2}, h_{p,3}, h_{p,4}\}$, C2P-Cache reads the corresponding rows from the Snapshot Matrix and performs a bitwise AND reduction:
\[
Cand_p = Row[h_{p,1}] \land Row[h_{p,2}] \land Row[h_{p,3}] \land Row[h_{p,4}],
\]
where each $Row[h_{p,i}]$ is an $N$-bit bitmap. A bit remains \textit{true} in $Cand_p$ only if all queried positions are \textit{true} in the corresponding column of the Snapshot Matrix. The resulting bitmap identifies the peer L1 caches that may contain the requested cache line.

\subsection{Parallel Snapshot Matrix Organization}
\label{sec:design-snapshot}

The Access Matrix encoding described above turns each L1 miss into a few row-indexed Snapshot Matrix lookups. As SMs may issue misses concurrently, the Snapshot Matrix must sustain high lookup bandwidth without becoming a bottleneck. C2P-Cache therefore organizes it as a banked and replicated structure that supports parallel row reads.

\subsubsection{Parallel Snapshot Matrix Organization}
\label{sec:design-snapshot-organization}

\begin{figure}[t]
    \centering
    \includegraphics[width=\linewidth]{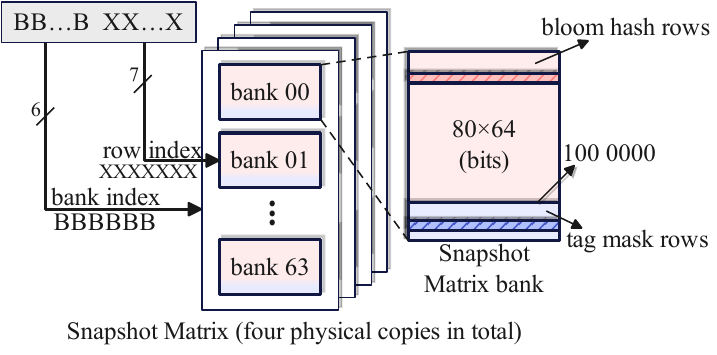}
    \caption{Parallel snapshot matrix organization.}
    \label{fig:snapshot_bank}
\end{figure}

The Snapshot Matrix is implemented using a banked organization with four physical copies in total. In the default configuration, the logical table contains $5120$ rows, each $64$-bit wide. C2P-Cache distributes these rows across $64$ banks, so that each bank stores $80$ rows.  Figure~\ref{fig:snapshot_bank} shows the physical organization. Each queried index is decoded into a bank identifier and a row offset. In the unified $13$-bit query format introduced in Section~\ref{sec:design-query-encoding}, the higher $6$ bits select the bank, while the remaining $7$ bits determine the row offset. This encoding unifies the addressing of tag-mask and BF-hash positions within the same Snapshot Matrix interface. Specifically, BF-hash indices use the form \texttt{\textbf{0}xxxxxx} and therefore map to rows 0--63 within each bank, whereas tag-mask indices use the form \texttt{\textbf{1}00xxxx} and therefore map to rows 64--79. As a result, each bank stores $64$ BF-hash rows and $16$ tag-mask rows, allowing both query types to share the same addressing path while remaining logically separated by row range.

This organization is designed to sustain high-concurrency row lookups on the L1 miss path. In the worst case, one miss from each of the $64$ SMs may enter C2P-Cache in a cycle, and each miss requires four Snapshot Matrix row lookups. The resulting demand is therefore $256$ row reads per cycle, which matches the provisioned bandwidth of $64$ banks with four physical copies per bank. Thus, when bank conflicts are modest, the design can sustain the worst-case lookup demand. Whether this parallelism can be realized in practice depends on bank-conflict behavior, which differs between the BF-hash and tag-mask regions. We therefore examine these two cases separately. For BF-hash positions, conflicts are naturally less severe because hashing tends to distribute queried positions evenly across banks. For tag-mask positions, C2P-Cache mitigates contention in two ways. First, the $1024$-bit tag-mask region is striped across all $64$ banks, so that each bank contains only $16$ tag-mask rows. Second, C2P-Cache applies bit reversal during tag-mask encoding to improve bank dispersion when nearby cache lines would otherwise map to the same few banks. Together, these mechanisms reduce contention in the tag-mask region while preserving the unified bank/index addressing path.

\subsubsection{Background Snapshot Refresh}
\label{sec:design-snapshot-refresh}
\begin{figure}[t]
    \centering
    \includegraphics[width=1\linewidth]{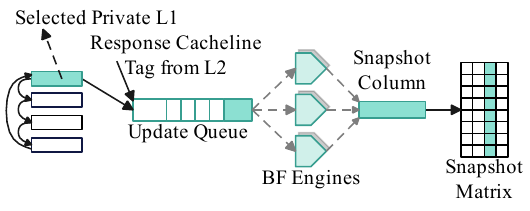}
    \caption{Refresh path of C2P-Cache.}
    \label{fig:SnapshotUpdate}
\end{figure}

Figure~\ref{fig:SnapshotUpdate} shows the background refresh path of C2P-Cache. Rather than tracking every private-L1 event exactly, C2P-Cache rebuilds each Snapshot Matrix column periodically from valid L1 tags, so replacements are reflected at the next rebuild.

During each period, C2P-Cache selects one L1 cache and extracts a subset of its valid tags into the Update Queue. As shown in Figure~\ref{fig:SnapshotUpdate}, these tags are processed by the BF Engines to rebuild the corresponding column of the Snapshot Matrix. The rebuild proceeds one SM at a time. Once an SM is selected, its target column is first cleared. The BF Engine encodings of all valid tags from the selected private L1 cache are then ORed into the Snapshot Column, which writes the updated contents to the corresponding column of the Snapshot Matrix every cycle until reconstruction is complete.

Cache lines returned from L2 use the same update path but follow a different update rule. When an L1 miss falls back to L2 and the requested line later returns, its tag is inserted into the Update Queue with the destination L1 identifier. This rule is decoupled from periodic refresh, although it uses the same BF Engines and update path. After BF Engine encoding, the generated positions are ORed into the corresponding Snapshot Matrix column without clearing that column.

A Snapshot Matrix column under rebuild remains accessible during L1-miss handling, so it may be incomplete until all queued tags from the selected SM have been processed. This may miss some valid remote-L1 reuse opportunities, but it does not affect correctness because misses that do not find a remote hit still follow the baseline L2 path. The update path shares the BF Engines with the miss path, and miss-side query generation always has priority. Snapshot staleness affects performance but not correctness. If a line is evicted after the last rebuild, its stale bits may remain set until the next rebuild and cause an unnecessary probe; if a line is filled before its column is rebuilt or incrementally updated, C2P-Cache may miss that reuse opportunity. In both cases, every remote hit still requires exact tag confirmation and unresolved requests fall back to L2.

\subsection{Remote Probing and Data Return}
\label{sec:design-probe}

\subsubsection{Remote Probing}
The output of Snapshot Matrix matching is a candidate bitmap over all SMs. A set bit indicates that the corresponding private L1 cache may contain the requested cache line, although this result is only approximate. C2P-Cache therefore performs a second-stage confirmation step by probing candidate peer L1 caches. For each miss, the requester SM is pruned from the candidate bitmap because the request has already been confirmed to miss in the local L1 cache. The remaining candidates are then probed serially in order of physical distance, so that nearer peer L1 caches are checked first. If a probe hits, C2P-Cache immediately terminates the remaining probes for that miss; otherwise, if all candidates miss, the request follows the baseline L2 path.

\subsubsection{Data Return}
To decouple candidate generation from exact remote confirmation, C2P-Cache places probe requests into a dedicated probe queue. Each request carries the requester identity, missed tag, and ordered candidate-SM list. Remote probes access the normal tag and data arrays of the candidate private L1 cache and may contend with local accesses from that SM. C2P-Cache does not wait indefinitely for remote confirmation: if the target L1 remains busy, the pending probe request aborts remaining probes and falls back to the baseline L2 path. When a probe hits, the requested line returns through the on-chip interconnect and fills into the requester's L1 using the baseline refill path. Thus, C2P-Cache changes only the data source, while preserving baseline refill and fallback behavior.

\subsection{Correctness and Design Scope}
\label{sec:design-correctness}

C2P-Cache is a read-side optimization that increases the chance of finding reusable data in peer private L1 caches before a miss is sent to L2. Because its query generation is approximate and the Snapshot Matrix is refreshed only in the background, C2P-Cache may occasionally introduce false positives or false negatives. These effects impact performance only, not correctness, because any unresolved request still follows the baseline L2 path.

A false positive occurs when the Result Matrix indicates that a peer L1 cache may contain the requested cache line even though it does not. In this case, C2P-Cache may issue an unnecessary remote probe; however, correctness is preserved because the data is accepted only after exact tag confirmation in the probed L1 cache. A false negative occurs when C2P-Cache fails to identify a peer L1 cache that actually holds the requested line. This inaccuracy arises from snapshot staleness rather than from the BF Engine encoding itself, since the Snapshot Matrix is refreshed periodically, and a column under rebuild remains queryable. Although such cases miss an otherwise exploitable remote-L1 reuse opportunity, the request still falls through to the baseline L2 path.

C2P-Cache changes only the handling of read misses. It does not replace any correctness-critical metadata structure in the memory hierarchy, nor does it act as a coherence directory. Instead, C2P-Cache is a performance-oriented mechanism that adds an opportunistic peer-L1 lookup path before the baseline L2 accesses.

\section{Experiments and Evaluation}

\subsection{Experimental Methodology}
\label{sec:eval-method}

\begin{table}[t]
\normalsize
\centering
\footnotesize
% \normalsize
\caption{Evaluation setup and default C2P-Cache parameters.}
\label{tab:eval_setup}
\setlength{\tabcolsep}{4pt}
\begin{tabular}{l p{0.62\linewidth}}
\toprule
\textbf{Parameter} & \textbf{Value} \\
\midrule
\multicolumn{2}{l}{\textit{(a) Baseline GPU configuration}} \\
GPU configuration & 64 SMs, 1.41\,GHz \\
SM organization & 8 SMs/cluster \\
Warp scheduler & GTO \cite{gto}, 4 schedulers/SM \\
Warp size/threads & 32 threads/warp, 2048 threads/SM \\
Registers & 65536 registers/SM \\
Private L1 cache & 64\,KB/SM, 4 sets, 32-way, 128\,B line, 20-cycle latency \\
Shared L2 cache & 128 sets, 16-way, 128\,B line, 200-cycle latency \\
Memory system & 20 memory partitions, 2 sub-partitions/partition \\
\midrule
\multicolumn{2}{l}{\textit{(b) Modeled latency assumptions (cycles)}} \\
ATA \cite{sharingl1-ATA} & aggregated tag array lookup: 7; cache line access: 14 \\
CCD \cite{sharingl1-CCD}& 2-bit counter lookup: 1; broadcast: 3; tag lookup: 7\\
RING \cite{sharingl1-RING}& hop latency: 2; copy-tag lookup: 7 \\
C2P-Cache & BF Engine: 2 \cite{BFlatencyFPGA,bffpga,BFFPGA2}; Snapshot Matrix lookup: 2; remote data return: 2; remote tag lookup: 7 \\
\midrule
\multicolumn{2}{l}{\textit{(c) Default C2P-Cache configuration}} \\
BF size ($m$) & 5120 bits \\
Tag-mask / BF-hash  & 1024 / 4096 bits \\
Encoding ($k$) & 1 tag-mask + 3 BF-hash \\
Snapshot organization & 64 banks, 4 copies per bank (1 original + 3 replicas) \\
BF Engine & 128 BF/tag-mask engines \\
Update transport bandwidth & 128\,B/cycle \cite{A100_white_paper}\\
\bottomrule
\end{tabular}
\renewcommand{\arraystretch}{1.0}
\end{table}

\subsubsection{Simulators}
We evaluate C2P-Cache using Accel-Sim \cite{accel-sim}, a cycle-level GPU simulation framework, and choose the baseline configuration from the simulator model and NVIDIA documentation~\cite{V100_white_paper,A100_white_paper,H100_white_paper,GB200_white_paper,cuda_programming_guide,ldgsts}. Unless otherwise specified, Table~\ref{tab:eval_setup} summarizes the baseline, latency assumptions, and default C2P-Cache parameters. We model the BF Engine, Snapshot Matrix lookup/update paths, remote probing, exact remote tag confirmation, target-L1 contention, and L2 fallback in Accel-Sim. Remote traffic uses a dedicated far-L1 network with separate request and response latency queues: selected candidates enter banked remote-request queues, then the target SM's finite probe queue; hit responses return through response queues, while misses continue to the next candidate or fall back to the baseline interconnect/L2 path when queues or target resources are unavailable. The RTL area study implements the corresponding query queues, arbitration, bitmap reduction, peer selection, fallback, and probe/return control. For power, we use AccelWattch \cite{accel-watch,gpuwattch}; for Snapshot Matrix SRAM, CACTI \cite{cacti}.

\subsubsection{Benchmarks}
We evaluate C2P-Cache using 24 GPU workloads drawn from ISPASS \cite{ISPASS}, Rodinia 3.1 \cite{rodinia}, Parboil \cite{stratton2012parboil}, PolyBench \cite{polybenchGPU}, and Pannotia \cite{pannotia}, as listed in Table~\ref{tab:benchmarks}. These workloads span a broad range along two key dimensions relevant to C2P-Cache: the degree of redundant access that creates opportunities for peer-L1 reuse, and the extent to which performance is sensitive to L2 cache latency.

\subsubsection{Compared Designs}
The baseline GPU does not include peer-L1 lookup support; L1 misses are served by the shared L2 through the memory hierarchy. We compare C2P-Cache against this baseline and representative L1 sharing mechanisms, including ATA \cite{sharingl1-ATA}, CCD \cite{sharingl1-CCD}, and RING \cite{sharingl1-RING}. All compared designs use the same baseline GPU configuration unless noted. For the main comparison, CCD and ATA use the cluster-based organizations described in their papers, because their broadcast mechanisms and aggregated structures constrain parallelism. We also sweep the sharing cluster from 8 to 64 SMs, where 64 SMs is chip-wide.

% \begin{table}[t]
% \centering
% \footnotesize
% \caption{Baseline GPU configuration used in our evaluation.}
% \label{tab:gpu_config}
% \setlength{\tabcolsep}{4pt}
% \begin{tabular}{l p{0.62\linewidth}}
% \toprule
% \textbf{Parameter} & \textbf{Value} \\
% \midrule
% GPU configuration & 64 SMs, 1.41\,GHz \\
% SM organization & 8 SMs/cluster \\
% Warp scheduler & GTO, 4 schedulers/SM, single issue per warp \\
% Warp size / threads & 32 threads/warp, 2048 threads/SM \\
% Registers & 65536 registers/SM \\
% Private L1 cache & 64\,KB/SM, 4 sets, 32-way, 128\,B line, 20-cycle latency \\
% Shared L2 cache & 128 sets, 16-way, 128\,B line, 200-cycle latency \\
% Memory system & 20 memory partitions, 2 sub-partitions/partition \\
% \bottomrule
% \end{tabular}
% \end{table}

\begin{table}[t]
\centering
\footnotesize
\caption{Benchmarks used in C2P-Cache.}
\label{tab:benchmarks}
\setlength{\tabcolsep}{4pt}
\begin{tabular}{lll@{\hspace{6pt}}lll}
\toprule
Benchmark & Suite & Abbr. & Benchmark & Suite & Abbr. \\
\midrule
BFS & ISPASS & BV & sgemm & Parboil & SG \\
LIB & ISPASS & LI & stencil & Parboil & ST \\
LPS & ISPASS & LP & 2DConvolution & PolyBench & 2D \\
RAY & ISPASS & RA & 3mm & PolyBench & 3M \\
b+tree & Rodinia 3.1 & B+ & atax & PolyBench & AT \\
dwt2d & Rodinia 3.1 & DW & bicg & PolyBench & BI \\
gaussian & Rodinia 3.1 & GA & gemm & PolyBench & GE \\
hotspot1 & Rodinia 3.1 & HO & gesummv & PolyBench & GS \\
lud & Rodinia 3.1 & LU & color\_max & Pannotia & CO \\
nn & Rodinia 3.1 & NN & fw\_block & Pannotia & FW \\
cutcp & Parboil & CU & mis & Pannotia & MI \\
mri & Parboil & MR & pagerank & Pannotia & PA \\
\bottomrule
\end{tabular}
\end{table}

% \subsection{Design Point and Timing Assumptions}
% \label{sec:eval-design-point}

\subsubsection{Timing Assumptions}
\label{sec:eval-timing}

% \begin{table}[t]
% \centering
% \footnotesize
% \caption{Modeled latency assumptions.}
% \label{tab:mechanism_latency}
% \setlength{\tabcolsep}{4pt}
% \begin{tabular}{l p{0.7\linewidth}}
% \toprule
% \textbf{Mechanism} & \textbf{Modeled latency assumptions} \\
% \midrule
% ATA & aggregated tag array lookup: 7 cycles; cacheline access: 14 cycles \\
% CCD & 2-bit counter lookup: 1 cycle; broadcast: 3 cycles; tag lookup: 7 cycles \\
% RING & hop latency: 2 cycles; copy-tag lookup: 7 cycles \\
% C2P-Cache & BF Engine: 2 cycle; Snapshot Matrix matching: 2 cycles; remote data return: 2 cycles; remote tag lookup: 7 cycles \\
% \bottomrule
% \end{tabular}
% \end{table}

Table~\ref{tab:eval_setup}(b) summarizes the modeled component latencies for the evaluated L1 cache-sharing mechanisms. These assumptions follow the cycle-level modeling style used by GPGPU-Sim/Accel-Sim GPU architecture studies~\cite{ISPASS,modern_gpu_modeling,accel-sim,accel-watch,gpuwattch} and the relative component latencies reported by prior L1-sharing mechanisms~\cite{sharingl1-ATA,sharingl1-CCD,sharingl1-RING,sharingl1-MeshL1sharing}. For communication components such as broadcast, hop latency, and remote data return, we use explicit cycle-level miss-path costs, consistent with remote-L1 mechanisms and NoC simulation studies~\cite{sharingl1-RING,sharingl1-MeshL1sharing,booksim_noc}. For remote tag confirmation, we conservatively model tag lookup as 7 cycles, motivated by CACTI-style cache timing estimates~\cite{cacti} that separate tag and data paths and suggest that tag-only access is shorter than full cache access. It is also consistent with CCD \cite{sharingl1-CCD}, where tag lookup is modeled as 1 cycle versus 3 cycles for cache access.

For C2P-Cache, we model the BF Engine as 2-cycle operations. The BF Engine latency is motivated by prior hardware-oriented Bloom-filter designs, which report 1-cycle hash generation and 2-cycle Bloom-filter lookup with optimized lightweight hashing~\cite{BFlatencyFPGA,bffpga,BFFPGA2,bfstructurecache}. To validate Snapshot Matrix timing, we use CACTI \cite{cacti} to model the per-bank SRAM arrays and find that one row can be read within one cycle. After accounting for post-read combination logic, including bitwise reduction over returned row bitmaps, we model the Snapshot Matrix lookup as 2 cycles. C2P-Cache runs after a local L1 miss, so it is off the L1-hit critical path. The 2-cycle remote-return latency is only the base hit-response cost; queueing, target-L1 contention, fallback, and response backpressure are dynamically modeled, represented in RTL control, and swept up to 32 cycles.

\subsubsection{BF Parameters}
\label{sec:eval-default-point}

% \begin{figure}[t]
%     \centering
%     \includegraphics[width=0.8\linewidth]{fig/evaluation/mkFPbig.pdf}
%     \caption{FP rate under different ($m$) and ($k$)}
%     \label{fig:bf_design_point}
% \end{figure}

C2P-Cache uses a default configuration chosen by balancing filtering effectiveness against miss-path overhead, rather than by minimizing false positives alone. 

% The false-positive rate of a Bloom filter is given by
\begin{equation}
P_{\mathrm{FP}}
% =
% \left(1-\left(1-\frac{1}{m}\right)^{kn}\right)^k
\approx
\left(1-e^{-kn/m}\right)^k ,
\label{eq:bloom-fp}
\end{equation}
According to Eq.~\eqref{eq:bloom-fp}~\cite{bloomfilter,bloomfilterFP}, we adopt a 5120-bit BF with \(k=4\), which achieves an approximately 1\% false-positive rate.

% Based on the equation, we choose a 5120-bit BF with \(k=4\), which achieves an approximately 1\% false-positive rate.

C2P-Cache further adopts a hybrid encoding with a 1024-bit tag-mask region and a 4096-bit BF-hash region, using one tag-mask position and three BF-hash positions per miss. This design keeps query width and Snapshot Matrix access cost low while preserving sufficient filtering effectiveness.

% \begin{table}[t]
% \centering
% \footnotesize
% \caption{Default C2P-Cache configuration.}
% \label{tab:bfcache_config}
% \setlength{\tabcolsep}{4pt}
% \begin{tabular}{l p{0.5\linewidth}}
% \toprule
% \textbf{Parameter} & \textbf{Value} \\
% \midrule
% BF size ($m$) & 5120 bits \\
% Tag-mask / BF-hash split & 1024 / 4096 bits \\
% Encoding & 1 tag-mask + 3 BF-hash \\
% Snapshot organization & 64 banks, 4 replicas/bank \\
% BF update buffer & 512 entries \\
% Update-path tag transport bandwidth & 128\,B/cycle \\
% \bottomrule
% \end{tabular}
% \end{table}

\subsection{Performance Evaluation}
\label{sec:eval-performance}

\begin{figure*}[t]
  \centering
  \includegraphics[width=\textwidth]{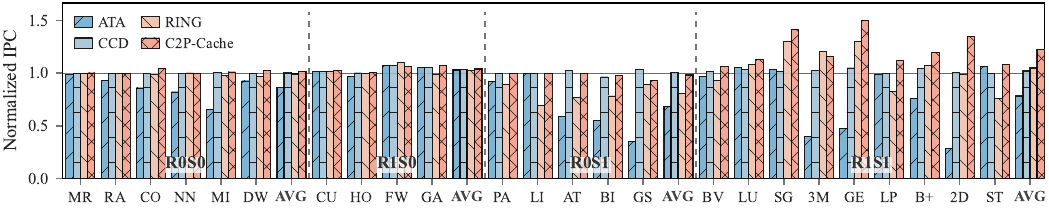}
  \caption{Normalized IPC across four workload groups.}
  \label{fig:ipc}
\end{figure*}

We compare C2P-Cache with a baseline GPU and three L1 cache sharing designs: ATA \cite{sharingl1-ATA}, CCD \cite{sharingl1-CCD}, and RING \cite{sharingl1-RING}. In the baseline GPU, each SM has a private L1 cache. ATA uses an aggregated tag array to handle all cache accesses, CCD enables predictor-guided cluster-level sharing, and RING forwards miss requests along a ring to probe remote caches. All results are normalized to the baseline. All designs use the configuration shown in Table \ref{tab:eval_setup}.

\subsubsection{Normalized IPC}

Figure~\ref{fig:ipc} reports normalized IPC for all evaluated workloads, grouped into four categories according to the characterization in Section~\ref{sec:motivation}: R0S0, R1S0, R0S1 and R1S1. 

We make three key observations. 
\textbf{First}, C2P-Cache is effective for R1S1 workloads, which are most amenable to L1 cache sharing, improving IPC by up to 49.7\% and by 23.5\% on average over the baseline and outperforming all prior sharing designs. It remains beneficial in R0S0 and R1S0, improving IPC by 1.4\% and 4.2\% on average, respectively. Although it incurs a 2.0\% IPC degradation in R0S1, where peer-L1 reuse is scarce despite high sensitivity to miss-path latency, this degradation is much smaller than ATA and RING, which reduce IPC by 31.7\% and 19.3\%, while CCD delivers only a slight 0.4\% IPC improvement. These results confirm that C2P-Cache's main benefit comes from R1S1-like workloads. 
\textbf{Second}, sharing scope has an impact on effectiveness. On R1S1, chip-wide C2P-Cache and RING outperform cluster-local CCD and ATA, indicating that broader visibility recovers additional reuse opportunities.
\textbf{Third}, pruning-based designs provide more consistent performance across workloads. By suppressing cache sharing when its benefit is unlikely to outweigh its cost, pruning reduces unnecessary overhead and interference. In contrast, ATA degrades on miss-intensive workloads such as \textit{3M}, \textit{2D}, and \textit{GE} due to the limited throughput of its aggregated tag-array path, whereas RING is less effective on workloads such as \textit{LI}, \textit{ST}, and \textit{LP}, where long-distance propagation and ring-network contention increase sharing overhead. Consequently, C2P-Cache show smaller performance variation across workload categories.

Overall, C2P-Cache combines broad visibility with pruning, capturing inter-SM reuse while avoiding many degradations of prior sharing designs.

\begin{figure}[t]
  \centering
  \includegraphics[width=\linewidth]{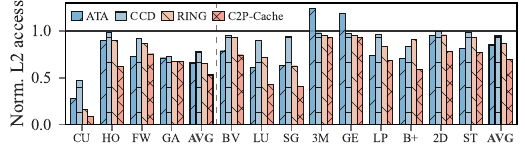}
  \caption{Normalized L2 accesses for R1S0 and R1S1.}
  \label{fig:l2access}
\end{figure}

\subsubsection{L2 Access Variation}
\label{sec:eval-l2access}

To evaluate the impact on the L2 cache, we measure L2 accesses normalized to the baseline. Figure~\ref{fig:l2access} reports the results for the high-redundancy R1S0 and R1S1 groups.

We make two observations. \textbf{First}, C2P-Cache achieves the largest L2-access reduction, with average normalized L2 accesses of 53.4\% for R1S0 and 69.8\% for R1S1. This benefit comes from its chip-wide visibility, unlike RING, whose serialized traversal is vulnerable to network congestion. \textbf{Second}, L2-access reduction correlates with IPC improvement, with ATA being a notable exception. Under high miss pressure, its aggregated tag array becomes a bottleneck, causing more requests to proceed directly to L2. As a result, \textit{3M} and \textit{GE} show increased L2 accesses in our implementation.

Overall, these results show that C2P-Cache reduces L2 accesses and translates such reduction into better performance gains.

\subsubsection{Prediction and Filtering Accuracy}
\label{sec:eval-tpfp}

% To evaluate pruning between CCD and C2P, we examine the quality of miss-time candidate generation.

\begin{figure}[t]
  \centering
  \includegraphics[width=\linewidth]{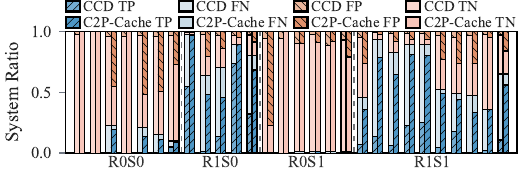}
  \caption{System-level TP/FN/FP/TN of CCD and C2P-Cache.}
  \label{fig:tpfp}
\end{figure}

To evaluate pruning effectiveness in CCD and C2P-Cache, we characterize miss-time candidate generation. Figure~\ref{fig:tpfp} reports the fractions of TP, FN, FP, and TN accesses for CCD and C2P-Cache, normalized to total L1 misses, so the four categories sum to 1 for each workload. TP corresponds to reusable L1 misses retained, FN to reusable L1 misses incorrectly pruned, FP to non-reusable L1 misses retained, and TN to non-reusable L1 misses correctly pruned. These metrics reflect system-level outcomes rather than the per-query BF FP rate in Eq.~\ref{eq:bloom-fp}.

We make three observations. \textbf{First}, C2P-Cache achieves a higher TP rate than CCD, especially for R1S0 and R1S1 workloads, indicating that chip-wide BF-based filtering recovers more useful peer-L1 hits. \textbf{Second}, C2P-Cache also incurs a higher FP rate, expected because it operates over a broader sharing scope and uses approximate filtering before exact probing. However, the TP increase is much larger than the FP increase. As a result, C2P-Cache delivers higher overall pruning effectiveness than CCD; for example, on R1S1, its combined TP+TN ratio reaches 76.5\%, compared with 43.4\% for CCD. \textbf{Third}, this TP--FP trade-off is consistent with Figure~\ref{fig:ipc}: C2P-Cache is most effective when reusable peer-L1 data is abundant, whereas FP cost becomes more visible when reuse is limited. However, this trend is not universal. For example, \textit{PA} exhibits relatively high FP activity without a clear IPC loss, indicating that FP overhead alone does not determine performance.

Overall, compared with CCD, C2P-Cache achieves more effective pruning by retaining more useful sharing opportunities, demonstrating that our design is more effective than a simple 2-bit counter.

% \tightenfloatspacing

\begin{figure}[t]
  \centering
  \includegraphics[width=\linewidth]{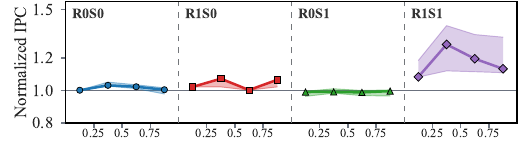}
  \caption{IPC vs. FP ratio.}
  \label{fig:fp_ipc}
\end{figure}

\subsubsection{False-Positive Impact on IPC}
\label{sec:eval-fp-ipc}

To evaluate false positives, we vary \(m\) and \(k\) to obtain different system-level FP ratios. Figure~\ref{fig:fp_ipc} bins all points by FP interval; the line shows the median IPC, and the shaded band covers the middle 50\% of points (25th--75th percentile). Because C2P-Cache also includes exact tag confirmation, candidate ordering, queues, target-L1 contention, and L2 fallback, FP ratio is not the only performance factor.

We make three observations. \textbf{First}, in R1S1, IPC first rises and then falls as FP increases: low-FP settings prune aggressively and also lose true positives, medium-FP settings expose more useful remote hits, and high-FP settings add enough useless probes to reduce performance. \textbf{Second}, R0S0, R1S0, and R0S1 are much less sensitive to FP changes because they either have less miss-path pressure or fewer useful remote hits. \textbf{Third}, the shaded band widens at higher FP ratios, showing that performance becomes more workload-dependent as queueing, target-L1 contention, and fallback interact with the TP--FP balance.

Overall, C2P-Cache must control FPs, but IPC is governed by the system-level TP--FP trade-off rather than by Bloom-filter FP probability alone.

\begin{figure}[t]
  \centering
  \includegraphics[width=\linewidth]{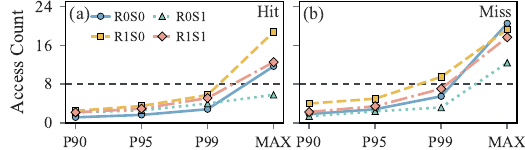}
  \caption{Peer-L1 access-count distribution.}
  \label{fig:worst_percentile}
\end{figure}

\subsubsection{L1 Cache Access Impact}
\label{sec:eval-accesscount}

To quantify extra L1-cache accesses induced by cache sharing, we measure the number of candidate L1 caches consulted by successful (Hit) and unsuccessful (Miss) probes. Figure~\ref{fig:worst_percentile} summarizes the peer-L1 access-count distribution and includes the ATA/CCD reference cost. P90, P95, P99, and max denote the 90th, 95th, 99th percentiles and maximum per-probe access counts; plotted values are averaged across benchmarks in each workload group. ATA's aggregated tag-array lookup and CCD's broadcast probing access every L1 cache in the target cluster, so both incur 8 peer-L1 lookups per remote sharing attempt. In contrast, RING accesses one remote L1 cache only after a hit has been verified using replicated cache tags, and avoids peer-L1 probing on misses.

We make two observations. \textbf{First}, despite supporting chip-wide sharing, C2P-Cache's P90 access count is only 2.01 on hits and 2.38 on misses, far below the ATA/CCD reference line. \textbf{Second}, the group-averaged tails remain modest: P95 stays below 3.5 for hits and 5.0 for misses, and P99 remains below 5.8 for hits and 9.5 for misses. Rare workload-level extremes, such as \textit{FW} requiring up to 63 checks before a hit, occur beyond the 99th percentile and are bounded by queue backpressure and L2 fallback.

Overall, C2P-Cache enables chip-wide sharing with modest typical L1 lookup overhead.

\subsection{Overhead}

\subsubsection{Power Overhead}
\label{sec:eval-power}

\begin{figure*}[ht]
  \centering
  \includegraphics[width=\linewidth]{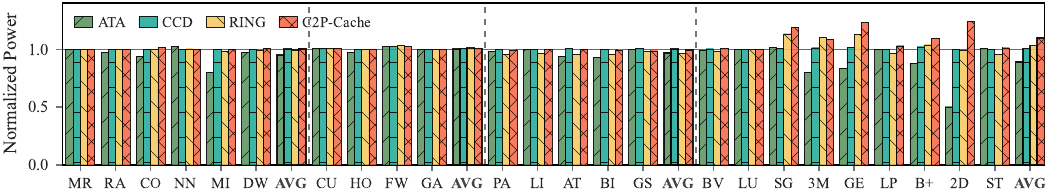}
  \caption{Normalized power of ATA \cite{sharingl1-ATA}, CCD \cite{sharingl1-CCD}, RING \cite{sharingl1-RING}, and C2P-Cache.}
  \label{fig:power}
\end{figure*}

We use AccelWattch~\cite{accel-watch,gpuwattch} to evaluate the power overhead of different sharing mechanisms. Figure~\ref{fig:power} reports the normalized power of ATA, CCD, RING, and C2P-Cache.

We make two observations. \textbf{First}, ATA lowers power on memory-intensive workloads by limiting each cluster to four L1 requests per cycle, but this bottleneck also reduces IPC; on R1S1, its performance per watt is only \(0.661/0.891=0.741\). \textbf{Second}, C2P-Cache increases R1S1 power by 9.8\%, but its larger IPC gain yields \(1.235/1.098=1.125\) performance per watt, above CCD (\(1.019/1.007=1.013\)) and RING (\(1.024/1.031=0.993\)). On R0S0 and R1S0, C2P-Cache changes performance per watt by +1.0\% and +3.6\%, respectively.

\subsubsection{Area Overhead}
\label{sec:eval-area}

% We estimate SRAM arrays with CACTI~\cite{cacti} at 32\,nm. To compare against ASAP7 logic, we report a 7\,nm-equivalent SRAM area scaled by $(7/32)^2$. We synthesize non-SRAM C2P logic with Yosys/ASAP7~\cite{yosys,yosys_nextpnr,asap7,asap7_stdcell}, following common early-stage area modeling practice~\cite{cacti,mcpat}. 
We estimate Snapshot Matrix SRAM with CACTI~\cite{cacti} at 32\,nm and report a 7\,nm-equivalent area scaled by $(7/32)^2$; non-SRAM C2P logic is synthesized with Yosys/ASAP7~\cite{yosys,yosys_nextpnr,asap7,asap7_stdcell}. Table~\ref{tab:c2p_area_overhead} separates capacity, SRAM area, BF-Engine cost, synthesized control logic, and placement estimates. The RTL includes BF/tag-mask engines, queues, Snapshot Matrix addressing, bitmap reduction, peer selection, fallback, and probe/return control. It excludes Snapshot Matrix bitcells, baseline cache arrays, SM pipelines, and baseline GPU NoC.

\begin{table}[t]
\centering
\small
\caption{C2P-Cache hardware overhead.}
\label{tab:c2p_area_overhead}
\setlength{\tabcolsep}{2.5pt}
\renewcommand{\arraystretch}{1.03}
\begin{tabular}{@{}p{0.25\columnwidth} p{0.34\columnwidth} p{0.33\columnwidth}@{}}
\toprule
\textbf{Component} & \textbf{Basis} & \textbf{Result} \\
\midrule
Capacity & Design config. & 40/160\,KB log./phys. \\
SRAM & CACTI~\cite{cacti} & 1.68 $\rightarrow$ 0.0804\,mm$^2$ \\
BF Engine & VCU118~\cite{VCU118} synth. & 268 LUTs/engine \\
Logic & Yosys/ASAP7~\cite{yosys,asap7} & 0.0682\,mm$^2$; 668K cells \\
Placement & OpenROAD~\cite{openroad} & 0.0691/0.0837\,mm$^2$ \\
Total & scaled SRAM + logic & 0.149\,mm$^2$ \\
\bottomrule
\end{tabular}
\end{table}

The Snapshot Matrix has 40\,KB logical capacity, or 0.98\% of aggregate private-L1 capacity; four-copy banking raises physical capacity to 160\,KB, comparable to RING's 128\,KB shadow-tag storage~\cite{sharingl1-RING}. CACTI reports 1.68\,mm$^2$ at 32\,nm, or 0.0804\,mm$^2$ after $(7/32)^2$ scaling. With this normalization, the synthesized non-SRAM logic adds 0.068\,mm$^2$, and the total scaled C2P metadata/control overhead is 0.149\,mm$^2$. A VCU118 synthesis of one BF Engine uses 268 LUTs~\cite{VCU118}, and the complete non-SRAM RTL maps to 667{,}980 cells, 75{,}648 flip-flops, zero inferred memories, and 0.06819\,mm$^2$ in ASAP7. OpenROAD/ASAP7 placement reports 0.069/0.084\,mm$^2$ detailed/global area~\cite{openroad}; this is a sanity check, not signoff PPA.

% The complete non-SRAM RTL maps to 667{,}980 cells, 75{,}648 flip-flops, and zero inferred memories. An OpenROAD/ASAP7 placement sanity check reports 0.06907\,mm$^2$ detailed-placement area and 0.08369\,mm$^2$ global-placement area~\cite{openroad}; this is not signoff PPA.

% The Snapshot Matrix has 40\,KB logical capacity, or 0.98\% of aggregate private-L1 capacity; four-copy banking raises physical capacity to 160\,KB. Our CACTI run uses a 32\,nm technology setting and reports 1.68\,mm$^2$ for this physical Snapshot Matrix. To compare this SRAM estimate with the ASAP7 logic numbers under a common nominal technology, we additionally apply a simple geometric area normalization by $(7/32)^2$, giving 0.0804\,mm$^2$ at a 7\,nm-equivalent scale. This is comparable in capacity to prior remote-L1 metadata overheads such as RING's 128\,KB shadow-tag storage~\cite{sharingl1-RING}. With this normalization, the synthesized non-SRAM logic adds 0.068\,mm$^2$, and the total scaled C2P metadata/control overhead is 0.149\,mm$^2$. A VCU118 synthesis of one BF Engine uses 268 LUTs~\cite{VCU118}, and the complete non-SRAM RTL maps to 667{,}980 cells, 75{,}648 flip-flops, zero inferred memories, and 0.06819\,mm$^2$ in ASAP7. An OpenROAD/ASAP7 placement sanity check reports 0.06907\,mm$^2$ detailed-placement area and 0.08369\,mm$^2$ global-placement area~\cite{openroad}; this is not signoff PPA, but checks that the synthesized control logic is not dominated by placement expansion.

\subsection{Design Sensitivity}
\label{sec:eval-design-sensitivity}

To evaluate sensitivity, we vary one parameter at a time from the default configuration: sharing scope, Snapshot Matrix matching latency, remote data-return latency, BF Engine latency/parallelism, Snapshot Matrix organization, BF parameters \(m\) and \(k\), and SM count. Results are normalized IPC relative to the baseline GPU.

\begin{figure}[t]
  \centering
  \includegraphics[width=\linewidth]{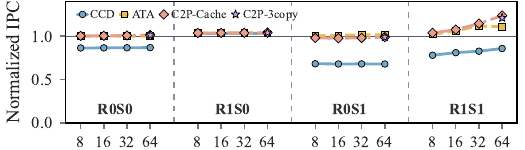}
  \caption{IPC vs. cluster size.}
  \label{fig:fair_cluster}
\end{figure}

\subsubsection{Sharing-Scope Sensitivity}
\label{sec:eval-fair}

% Figure~\ref{fig:fair_cluster} varies the sharing scope from 8 to 64 SMs; 64 SMs is chip-wide on our GPU. To enable iso-storage comparison, we include a C2P configuration with 3 copies, which uses 120\,KB of Snapshot Matrix storage. For CCD and ATA, we extend their tag-copy structures up to 128\,KB to match the same iso-storage budget, and allow them to bypass sharing when the shared queue is congested. This ensures that all designs are evaluated under comparable metadata storage constraints.

% We make three observations. \textbf{First}, under iso-storage conditions, C2P-Cache remains consistently best, reaching 21.2\% average IPC improvement at 64 SMs and outperforming chip-wide ATA and CCD. \textbf{Second}, CCD improves with scope but remains limited by broadcast traffic, staying 14.2\% below baseline at 64 SMs. \textbf{Third}, ATA improves up to 32 SMs but drops from 11.6\% to 10.8\% above baseline at 64 SMs due to centralized lookup contention, even with increased tag-copy capacity.

% Overall, under the same storage budget, chip-wide visibility alone is insufficient; scalable pruning is required for performance gains.

Figure~\ref{fig:fair_cluster} varies the sharing scope from 8 to 64 SMs; 64 SMs is chip-wide on our GPU. To enable iso-storage comparison, we include a C2P configuration with 3 copies, which uses 120\,KB Snapshot Matrix storage. For CCD and ATA, we extend their tag-copy structures to 128\,KB to match the same iso-storage budget, and allow them to bypass sharing when the shared queue is congested. This ensures all designs are evaluated under comparable metadata storage and sharing-scope constraints.

We make three observations. \textbf{First}, under iso-storage conditions, C2P-Cache remains consistently best, reaching 21.2\% average IPC improvement at 64 SMs and outperforming chip-wide ATA and CCD. \textbf{Second}, CCD improves with scope but remains limited by broadcast traffic, staying 14.2\% below baseline at 64 SMs. \textbf{Third}, ATA improves up to 32 SMs but drops from 11.6\% to 10.8\% above baseline at 64 SMs due to centralized lookup contention, even with increased tag-copy capacity.

Overall, chip-wide visibility alone is insufficient; scalable pruning is required for performance gains. We ensure matched storage and sharing-scope constraints for a symmetric comparison.

\subsubsection{Impact of Snapshot Matrix Matching Latency}
\label{sec:eval-mm-latency}

\begin{figure}[t]
  \centering
  \includegraphics[width=\linewidth]{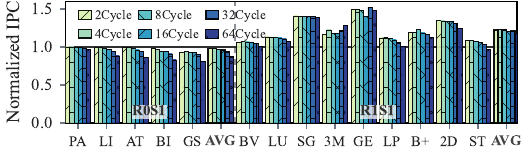}
  \caption{Snapshot-Matrix latency sensitivity.}
  \label{fig:mm_latency}
\end{figure}

We vary the Snapshot Matrix matching latency from 2 to 64 cycles. Figure~\ref{fig:mm_latency} reports R0S1 and R1S1, the two groups most sensitive to miss-path latency. R0S0 and R1S0 are omitted from the figure because their group averages remain close to the baseline even at 64 cycles.

We make two observations. \textbf{First}, C2P-Cache remains effective on R1S1 workloads even under longer matching latencies, with average IPC improvement decreasing from 23.5\% at 2 cycles to 17.8\% at 64 cycles. \textbf{Second}, R0S1 suffers the largest loss: its group-average IPC drops from 98.0\% to 87.0\%, because these workloads have limited reusable peer-L1 hits but remain sensitive to extra miss-path latency.

Overall, these results show that C2P-Cache is robust to Snapshot Matrix matching latency, suggesting that fewer Snapshot Matrix banks or replicas may incur limited performance loss.

\begin{figure}[t]
  \centering
  \includegraphics[width=\linewidth]{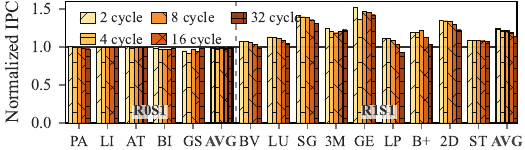}
  \caption{Remote data-return latency sensitivity.}
  \label{fig:remote_latency}
\end{figure}

\subsubsection{Impact of Remote Data-Return Latency}
\label{sec:eval-remote-latency}

To test whether C2P-Cache depends on an optimistic remote-return assumption, Figure~\ref{fig:remote_latency} varies remote data-return latency from 2 to 32 cycles. This latency is the base data-response cost after a remote tag hit; queueing, contention, and fallback are modeled dynamically.

We make two observations. \textbf{First}, the target R1S1 group remains effective even with much longer return latency. Its average IPC improvement decreases from 23.5\% at 2 cycles to 14.0\% at 32 cycles. \textbf{Second}, the impact depends on whether a workload has enough useful remote hits to amortize the extra miss-path cost. R1S0 remains 1.8\% above baseline at 32 cycles, whereas R0S0 and R0S1 remain close to baseline or slightly below it.

Overall, C2P-Cache's main benefit persists under longer remote-return latency, although workloads without useful peer-L1 reuse should avoid paying that cost.

\subsubsection{Impact of BF Encoding Latency and BF Parallelism}
\label{sec:eval-bf-lat-par}

\begin{figure}[t]
  \centering
  \includegraphics[width=\linewidth]{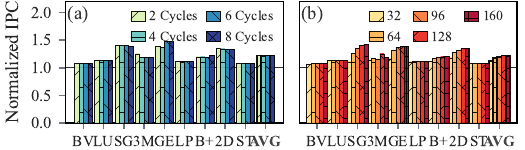}
  \caption{BF-Engine sensitivity: (a) latency and (b) count.}
  \label{fig:bf_lat_num}
\end{figure}
Figure~\ref{fig:bf_lat_num} examines BF Engine encoding latency and BF Engine count.

We make two observations. \textbf{First}, C2P-Cache is largely insensitive to moderate BF Engine encoding latency: increasing latency from 2 to 8 cycles changes R1S1 group-average IPC only marginally, from 23.5\% to 23.4\%. \textbf{Second}, increasing the number of BF Engines improves performance with diminishing returns. The group-average normalized IPC improvement increases from 12.8\% with 32 BF Engines to 18.8\%, 20.8\%, 23.5\%, and 22.8\% with 64, 96, 128, and 160 BF Engines. This trend suggests that scaling from 32 to 128 BF Engines helps absorb periods of high L1 miss pressure, while 160 BF Engines leave performance nearly unchanged; therefore, 128 engines are a practical default point.

Overall, C2P-Cache tolerates moderate BF Engine encoding latency and benefits from additional BF Engines.

\begin{figure}[!t]
  \centering
  \includegraphics[width=\linewidth]{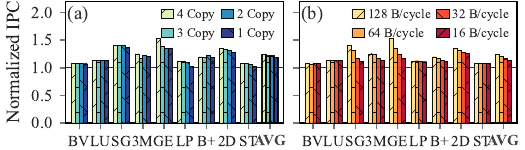}
  \caption{Snapshot-Matrix sensitivity: (a) copies, (b) BW.}
  \label{fig:snapshot_refresh_sens}
\end{figure}

\subsubsection{Impact of Snapshot Matrix Parameters}
\label{sec:eval-snapshot-refresh-sens}

Figure~\ref{fig:snapshot_refresh_sens} evaluates two Snapshot Matrix parameters on R1S1 workloads: physical Snapshot Matrix copies and update transport bandwidth. Reducing copies lowers lookup bandwidth, while reducing update bandwidth makes metadata staler.

We make two observations. \textbf{First}, C2P-Cache remains effective even with fewer Snapshot Matrix copies. The average IPC improvement is 23.5\% with four copies, and remains 21.2\%, 21.0\%, and 18.3\% with three, two, and one copy, respectively. Four copies use 160\,KB physical Snapshot Matrix storage, while three copies use 120\,KB and still retain most of the benefit; the fourth copy mainly protects extreme lookup bursts. \textbf{Second}, lower update bandwidth reduces performance gradually: the average IPC improvement remains 20.0\%, 15.6\%, and 13.4\% at 64, 32, and 16\,B/cycle.

Overall, C2P-Cache remains robust under reduced Snapshot Matrix capacity and update bandwidth, with performance mainly scaling with available remote-hit opportunities.

\subsubsection{Impact of Bloom-Filter Parameters}
\label{sec:eval-mk}

\begin{figure}[t]
  \centering
  \includegraphics[width=\linewidth]{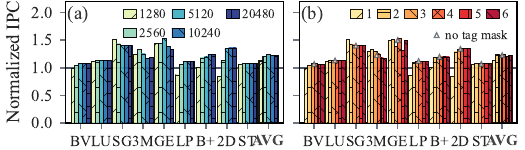}
  \caption{Sensitivity of BF parameters: (a) \(m\) and (b) \(k\).}
  \label{fig:mk}
\end{figure}

To evaluate sensitivity to BF parameters \(m\) and \(k\), Figure~\ref{fig:mk} reports normalized IPC of R1S1 workloads. In Figure~\ref{fig:mk}(a), the tag-mask portion is fixed at 1024 bits, and remaining storage is allocated to BF-hash regions with adjusted address mapping. In Figure~\ref{fig:mk}(b), \(k=1\) uses only the tag-mask, whereas \(k>1\) uses one tag-mask and \(k-1\) BF-hashes, generated as in the main design.

We make four observations. \textbf{First}, increasing \(m\) improves performance when \(m\) is small, but the benefit quickly saturates: the group-average IPC improvement rises from 12.0\% at \(m=1280\) to 23.5\% at \(m=5120\), with little gain at 10240 or 20480. \textbf{Second}, when \(k=1\), C2P-Cache achieves a 13.1\% average IPC improvement. Although this is lower than with additional BF-hashes, it still exceeds CCD's 1.9\% improvement, indicating that chip-wide visibility alone is beneficial, while tag-mask-only pruning leaves room for improvement. Once \(k \ge 2\), performance becomes insensitive to \(k\). We also evaluate a no-tag-mask variant that uses four BF-hash positions instead of the default one tag-mask plus three BF-hash positions. Its average improvement is 22.4\%, slightly below the default 23.5\%, showing that the tag-mask provides useful low-cost discrimination rather than merely replacing one BF hash. \textbf{Third}, the effects of \(m\) and \(k\) are not monotonic across workloads because the effective number of inserted elements \(n\), a key determinant of false-positive rate, changes dynamically during execution. \textbf{Fourth}, SG shows an anomalous trend in both Figure~\ref{fig:mk}(a) and Figure~\ref{fig:mk}(b), which we attribute to BF-hashes enabling more sharing opportunities but also introducing a small increase in bank conflicts.

Overall, C2P-Cache is robust to many BF parameter settings: performance improves as \(m\) increases from a small value but quickly saturates, while the design remains largely insensitive to \(k\) once BF-hashes are included. The default tag-mask plus three-BF encoding slightly outperforms the pure four-BF variant while using a simple lower-tag-bit fingerprint, so we keep the tag-mask to distinguish nearby cache lines and reduce hash-generation cost.

\subsubsection{Impact of SM Number}
\begin{figure}[t]
  \centering
  \includegraphics[width=\linewidth]{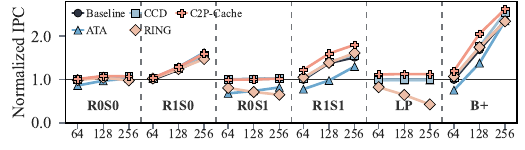}
  \caption{Sensitivity of SM number.}
  \label{fig:scalable}
\end{figure}

To evaluate scalability, Figure~\ref{fig:scalable} reports average IPC of the four workload groups at three SM scales, along with representative applications, \textit{LP} and \textit{B+}. We use a strong-scaling setup, where problem size is fixed as SM count increases.

We make three observations. \textbf{First}, cluster-based CCD and ATA are largely insensitive to scaling. Their normalized IPC follows a trend similar to the baseline because their sharing scope remains confined within each cluster and is not directly affected by chip-level scaling. \textbf{Second}, RING still exhibits unstable scalability due to network congestion. It degrades on R0S1 workloads and in some R1S1 cases. To illustrate this behavior, we highlight \textit{LP} and \textit{B+}, on which RING exhibits sharply different scaling trends. \textbf{Third}, C2P-Cache consistently delivers strong performance across all SM scales and workload groups.

Overall, the scaling results show that C2P-Cache's chip-wide pruning avoids the serialized discovery and centralized-structure bottlenecks that limit prior sharing designs as SM count increases.

\section{Discussion}
\label{sec:discussion}
\subsection{Workload Characteristics and R1S1 Reuse}
% \noindent\textbf{Workload characteristics and R1S1 reuse.}
% R1S1-like reuse stems from cross-SM redundancy in distributed GPU execution, where tiled kernels can map neighboring thread blocks to different SMs that read overlapping input regions. This pattern appears in GEMM, CNNs, Transformers, graph/tree traversal, stencils, and producer-consumer kernels, and recent inference studies report cross-layer and KV-cache reuse~\cite{ISPASS,rodinia,stratton2012parboil,polybenchGPU,pannotia,wang2020crosslayer,neo2024,kvcache2026}. These patterns are widely used in AI and HPC workloads, so R1S1-like behavior is not merely benchmark-specific. We do not claim that every workload is R1S1; rather, the classification separates the target cases from workloads where pruning and fallback must bound overhead. C2P-Cache therefore targets the cases where reuse and L2-latency sensitivity coincide, while keeping the cost small on workloads with little useful remote-L1 reuse.

R1S1-like reuse stems from cross-SM redundancy in distributed GPU execution, where tiled kernels map neighboring thread blocks to different SMs that access overlapping input regions. This pattern appears in GEMM, CNNs, Transformers, graph/tree traversal, stencils, and producer-consumer kernels, and is also observed in recent inference studies with cross-layer and KV-cache reuse~\cite{ISPASS,rodinia,stratton2012parboil,polybenchGPU,pannotia,wang2020crosslayer,neo2024,kvcache2026}. These patterns are widely used in AI and HPC workloads, so R1S1-like behavior is not benchmark-specific. We do not claim that all workloads exhibit R1S1; the classification separates target cases from workloads where pruning and fallback bound overhead. C2P-Cache targets cases where reuse and L2-latency sensitivity coincide while keeping overhead small otherwise.

\subsection{Relationship to Distributed Shared Memory}
% \noindent\textbf{Relationship to distributed shared memory.}
Recent GPUs support cooperative-cluster distributed shared memory~\cite{H100_white_paper,GB200_white_paper,cuda_programming_guide}, and software-managed shared memory/scratchpads expose explicit locality~\cite{sharedmemorysoftware1,sharedmemorysoftware2,stash,osm}. C2P-Cache instead targets implicit global-memory line reuse already present in private L1s, requires no program changes, and preserves the baseline L2 path. It can therefore help irregular, producer-consumer, or tiled global-memory reuse that is not always captured by managed shared-memory regions. Thus, the mechanisms are complementary.

\subsection{Bloom-filter Snapshot Design}
% \noindent\textbf{Bloom-filter snapshot design.}
A conventional insertion-only BF accumulates stale bits after L1 evictions~\cite{bloomfilter,bloomfilterFP}; C2P-Cache bounds this effect by rebuilding snapshots from valid L1 tags and inserting newly returned L2 lines. Deletable alternatives such as counting BF~\cite{countBF,bloom_finger_Sal}, cuckoo ~\cite{CuckooFilter}, and vector quotient filters~\cite{vqBF} require additional update metadata and eviction handling on the latency-sensitive path. C2P-Cache instead keeps the snapshot lightweight and off the L1-hit path: correctness is preserved via exact tag verification and L2 fallback.

% A conventional insertion-only Bloom filter accumulates stale bits after L1 evictions~\cite{bloomfilter,bloomfilterFP}; C2P-Cache bounds this by rebuilding snapshots from valid L1 tags and inserting newly returned L2 lines. Deletable alternatives such as counting Bloom filters~\cite{countBF,bloom_finger_Sal}, cuckoo filters~\cite{CuckooFilter}, and vector quotient filters~\cite{vqBF} require extra update metadata and eviction handling on the latency-critical path. C2P-Cache instead keeps the snapshot lightweight and off the L1-hit path: correctness is preserved via exact tag verification and L2 fallback.

\subsection{Security and Isolation}

The Snapshot Matrix stores lossy metadata rather than values or tags, but reflects approximate cache residency and is treated as microarchitectural state~\cite{gpu_side_channel_multigpu,gpu_timing_sidechannel}. It does not expose exact cache contents or allow inversion, but prior work shows cache-state and access-pattern side channels in GPUs~\cite{flush_reload,gpu_cache_attack}. Partitioning or flushing columns at context or kernel boundaries~\cite{mig_star} can reduce reuse without affecting correctness. We leave full side-channel analysis to future work.

% The Snapshot Matrix stores lossy BF/tag-mask metadata rather than data values or exact tags, but reflects approximate cache residency and is treated as microarchitectural state~\cite{gpu_side_channel_multigpu,gpu_timing_sidechannel}. It does not expose exact cache contents or allow inversion, but prior work shows cache-state and access-pattern side channels in GPUs~\cite{flush_reload,gpu_cache_attack}. Partitioning or flushing columns at context or kernel boundaries~\cite{mig_star} reduces reuse without affecting correctness. We leave full side-channel analysis to future work.

\section{Related Work}

Prior work has shown that private GPU L1 caches may hold duplicated data across SMs \cite{l1cacheredundancy1,l1cacheredundancy2,sharingl1-RING,sharingl1-colab}, creating redundant accesses to lower levels of the memory hierarchy while exposing opportunities for inter-SM reuse. Existing solutions follow two directions \cite{sharingl1-ATA}. One direction restructures the L1 organization to reduce replication or coordinate placement across cores \cite{sharingl1-L1.5Dcache,l1cacheredundancy2,sharingl1-decouplel1,sharingl1-two-levelshareing,scheduling4}. The other direction preserves private L1 caches and adds mechanisms to discover reusable data in peer L1 caches through sharing, lookup, redirection, or probing support \cite{sharingl1-ATA,sharingl1-CCD,sharingl1-Collaborative-Coalescing,sharingl1-colab,sharingl1-IntergroupCacheCooperation,sharingl1-MeshL1sharing,sharingl1-Pseudo-Cache,sharingl1-RING}. C2P-Cache belongs to the latter category.

Within this category, GPU L1 sharing techniques can be broadly grouped into three styles. Traversal-based schemes propagate requests among peer L1 caches to discover remote hits through distributed probing~\cite{sharingl1-RING}. Predictor-assisted schemes use reuse prediction or selective probing to determine when a request should be redirected to peer SMs~\cite{sharingl1-MeshL1sharing,sharingl1-CCD}. Metadata-assisted schemes rely on additional ownership, redirection, or aggregated-tag structures to determine whether a request should be served locally, by a peer L1, or by L2~\cite{sharingl1-ATA,sharingl1-colab,sharingl1-IntergroupCacheCooperation,sharingl1-Pseudo-Cache,sharingl1-Collaborative-Coalescing}. Although these designs differ in organization, they all require additional lookup, prediction, metadata, or probing support beyond the baseline private-L1 access path.

% As a result, their scalability is limited because each design makes a different trade-off among sharing scope, interference with normal accesses, and lookup concurrency. 

Table~\ref{tab:sharing_scheme_compare} summarizes three representative sharing designs and C2P-Cache in terms of native chip-wide sharing, interference to normal accesses, lookup concurrency, tag replication, and whether remote-hit discovery is exact or approximate. We select CCD~\cite{sharingl1-CCD}, ATA~\cite{sharingl1-ATA}, and Ring~\cite{sharingl1-RING} as representatives because they capture three design points: predictor-assisted selective sharing, metadata-assisted exact discovery through aggregated tags, and traversal-based distributed probing.

% Instead, it uses approximate candidate pruning followed by selective exact confirmation, trading a small amount of inaccuracy for broader visibility, lower interference, higher lookup concurrency, and no tag-replication overhead.

Bloom filters \cite{bloomfilter,bloomfilterFP} have long been used in cache architecture as compact, approximate metadata for filtering expensive operations. Representative applications include cache-miss prediction \cite{BFarchcachemiss}, NUCA cache search \cite{bfNUCA}, L2 cache miss prediction \cite{Morpheus}, cache prefetching \cite{bfarchcacheprefetching}, cache partitioning \cite{BF-arch1-cache-partitioning}, the Tagless Coherence Directory \cite{BFarchL2tag}, coherence-directory compression \cite{BF-arch-coherence}, cache-pollution filtering \cite{bfarchcachepollution}, synonym lookup \cite{bfarchvirtual_cache_synonym_lookup}, and cache-summary sharing \cite{bfcachesharing_network}. C2P-Cache instead uses BF snapshots as a cache-structure-aware, high-concurrency pruning path for L1 sharing.

\begin{table}[t]
\centering
\caption{Comparison of GPU L1 cache sharing schemes.}
\label{tab:sharing_scheme_compare}
\footnotesize
\setlength{\tabcolsep}{3.5pt}
\renewcommand{\arraystretch}{1.08}
\begin{tabular}{lcccc}
\toprule
\textbf{Metric} & \textbf{CCD \cite{sharingl1-CCD}} & \textbf{ATA \cite{sharingl1-ATA}} & \textbf{RING \cite{sharingl1-RING}} & \textbf{C2P-Cache} \\
\midrule
Native Chip-wide
& \textcolor{red}{\ding{55}} 
& \textcolor{red}{\ding{55}} 
& \textcolor{green!50!black}{\ding{51}} 
& \textcolor{green!50!black}{\ding{51}} \\

Low Interference
& \textcolor{red}{\ding{55}} 
& \textcolor{red}{\ding{55}} 
& \textcolor{green!50!black}{\ding{51}} 
& \textcolor{green!50!black}{\ding{51}} \\

High Concurrency
& \textcolor{red}{\ding{55}} 
& \textcolor{red}{\ding{55}} 
& \textcolor{red}{\ding{55}} 
& \textcolor{green!50!black}{\ding{51}} \\

No Tag Replication
& \textcolor{green!50!black}{\ding{51}} 
& \textcolor{green!50!black}{\ding{51}} 
& \textcolor{red}{\ding{55}} 
& \textcolor{green!50!black}{\ding{51}} \\

Exact Discovery
& \textcolor{red}{\ding{55}} 
& \textcolor{green!50!black}{\ding{51}} 
& \textcolor{green!50!black}{\ding{51}} 
& \textcolor{red}{\ding{55}} \\
\bottomrule
\end{tabular}
\end{table}

% \section{Conclusion}

\section{Conclusion}

We presented C2P-Cache, a scalable GPU L1 cache sharing mechanism that reduces redundant L2 accesses by transforming remote-L1 discovery from exact chip-wide search into lightweight candidate pruning followed by selective exact confirmation. C2P-Cache maintains BF-based L1-tag snapshots, provides chip-wide reuse visibility, and supports high-concurrency miss-time filtering through a banked and replicated Snapshot Matrix. Across diverse GPU workloads, C2P-Cache outperforms representative prior sharing designs, improving IPC by up to 49.7\% and by 23.5\% on average for applications with abundant remote-L1 reuse and high sensitivity to the L2 access path. More broadly, our results show that scalable GPU L1 sharing does not require exact chip-wide visibility on the miss path; it requires only enough information to eliminate most impossible locations before exact probing.

%%%%%%% -- PAPER CONTENT ENDS -- %%%%%%%%

%%
%% The next two lines define the bibliography style to be used, and
%% the bibliography file.
\setlength{\bibsep}{0pt}
\bibliographystyle{ACM-Reference-Format}
\bibliography{sample-base}

@manual{cuda_programming_guide,
  author = {{NVIDIA}},
  title = {{CUDA C++ Programming Guide}},
  organization = {{NVIDIA}},
  year = {2026},
  url = {https://docs.nvidia.com/cuda/cuda-c-programming-guide/}
}

@manual{mig_star,
  author = {{NVIDIA}},
  title = {{NVIDIA Multi-Instance GPU User Guide}},
  organization = {{NVIDIA}},
  year = {2026},
  url = {https://docs.nvidia.com/datacenter/tesla/mig-user-guide/}
}

@misc{yosys,
  title={Yosys-a free verilog synthesis suite},
  author={Wolf, Clifford and Glaser, Johann and Kepler, Johannes},
  booktitle={Proceedings of the 21st Austrian Workshop on Microelectronics (Austrochip)},
  volume={97},
  pages={1--6},
  year={2013},
  url = {https://yosyshq.net/yosys/},
}

@misc{yosys_nextpnr,
  author={Shah, David and Hung, Eddie and Wolf, Clifford and Bazanski, Serge and Gisselquist, Dan and Milanovic, Miodrag},
  booktitle={2019 IEEE 27th Annual International Symposium on Field-Programmable Custom Computing Machines (FCCM)}, 
  title={Yosys+nextpnr: An Open Source Framework from Verilog to Bitstream for Commercial FPGAs}, 
  year={2019},
  volume={},
  number={},
  pages={1-4},
  doi={10.1109/FCCM.2019.00010}}

@misc{openroad,
author = {Ajayi, Tutu and Chhabria, Vidya A. and Foga\c{c}a, Mateus and Hashemi, Soheil and Hosny, Abdelrahman and Kahng, Andrew B. and Kim, Minsoo and Lee, Jeongsup and Mallappa, Uday and Neseem, Marina and Pradipta, Geraldo and Reda, Sherief and Saligane, Mehdi and Sapatnekar, Sachin S. and Sechen, Carl and Shalan, Mohamed and Swartz, William and Wang, Lutong and Wang, Zhehong and Woo, Mingyu and Xu, Bangqi},
title = {Toward an Open-Source Digital Flow: First Learnings from the OpenROAD Project},
year = {2019},
isbn = {9781450367257},
publisher = {Association for Computing Machinery},
address = {New York, NY, USA},
url = {https://doi.org/10.1145/3316781.3326334},
doi = {10.1145/3316781.3326334},
booktitle = {Proceedings of the 56th Annual Design Automation Conference 2019},
articleno = {76},
numpages = {4},
location = {Las Vegas, NV, USA},
series = {DAC '19}
}

@article{asap7,
title = {ASAP7: A 7-nm finFET predictive process design kit},
journal = {Microelectronics Journal},
volume = {53},
pages = {105-115},
year = {2016},
issn = {1879-2391},
doi = {https://doi.org/10.1016/j.mejo.2016.04.006},
url = {https://www.sciencedirect.com/science/article/pii/S002626921630026X},
author = {Lawrence T. Clark and Vinay Vashishtha and Lucian Shifren and Aditya Gujja and Saurabh Sinha and Brian Cline and Chandarasekaran Ramamurthy and Greg Yeric}
}

@misc{asap7_stdcell,
  author={Xu, Xiaoqing and Shah, Nishi and Evans, Andrew and Sinha, Saurabh and Cline, Brian and Yeric, Greg},
  booktitle={2017 IEEE/ACM International Conference on Computer-Aided Design (ICCAD)}, 
  title={Standard cell library design and optimization methodology for ASAP7 PDK: (Invited paper)}, 
  year={2017},
  volume={},
  number={},
  pages={999-1004},
  doi={10.1109/ICCAD.2017.8203890}}

@inproceedings{modern_gpu_modeling,
  author = {Jain, Akshay and Khairy, Mahmoud and Ziabari, Amirsaman and Rogers, Timothy G. and Aamodt, Tor M.},
  title = {Exploring Modern {GPU} Memory System Design Challenges through Accurate Modeling},
  booktitle = {2019 IEEE International Symposium on Performance Analysis of Systems and Software},
  year = {2019},
  pages = {113--124},
  doi = {https://arxiv.org/abs/1810.07269}
}

@inproceedings{medic_gpu_memory,
  author={Ausavarungnirun, Rachata and Ghose, Saugata and Kayiran, Onur and Loh, Gabriel H. and Das, Chita R. and Kandemir, Mahmut T. and Mutlu, Onur},
  booktitle={2015 International Conference on Parallel Architecture and Compilation (PACT)}, 
  title={Exploiting Inter-Warp Heterogeneity to Improve GPGPU Performance}, 
  year={2015},
  volume={},
  number={},
  pages={25-38},
  doi={10.1109/PACT.2015.38}
}

@misc{gpu_side_channel_multigpu,
  author = {Dutta, Sankha Baran and Naghibijouybari, Hoda and Gupta, Arjun and Abu-Ghazaleh, Nael and Marquez, Andres and Barker, Kevin},
  title = {Spy in the GPU-box: Covert and Side Channel Attacks on Multi-GPU Systems},
  year = {2022},
  howpublished = {arXiv preprint arXiv:2203.15981},
  url = {https://arxiv.org/abs/2203.15981}
}

@inproceedings{gpu_timing_sidechannel,
  author = {Naghibijouybari, Hoda and Neupane, Ajaya and Qian, Zhiyun and Abu-Ghazaleh, Nael},
  title = {Rendered Insecure: GPU Side Channel Attacks are Practical},
  booktitle = {Proceedings of the 2018 ACM SIGSAC Conference on Computer and Communications Security},
  publisher = {Association for Computing Machinery},
  address = {New York, NY, USA},
  pages = {2139--2153},
  year = {2018},
  doi = {10.1145/3243734.3243831}
}

@ARTICLE{countBF,
  author={Li Fan and Pei Cao and Almeida, J. and Broder, A.Z.},
  journal={IEEE/ACM Transactions on Networking}, 
  title={Summary cache: a scalable wide-area Web cache sharing protocol}, 
  year={2000},
  volume={8},
  number={3},
  pages={281-293},
  doi={10.1109/90.851975}}

@inproceedings{CuckooFilter,
author = {Fan, Bin and Andersen, Dave G. and Kaminsky, Michael and Mitzenmacher, Michael D.},
title = {Cuckoo Filter: Practically Better Than Bloom},
year = {2014},
isbn = {9781450332798},
publisher = {Association for Computing Machinery},
address = {New York, NY, USA},
url = {https://doi.org/10.1145/2674005.2674994},
doi = {10.1145/2674005.2674994},
booktitle = {Proceedings of the 10th ACM International on Conference on Emerging Networking Experiments and Technologies},
pages = {75–88},
numpages = {14},
location = {Sydney, Australia},
series = {CoNEXT '14}
}

@inproceedings{vqBF,
author = {Pandey, Prashant and Conway, Alex and Durie, Joe and Bender, Michael A. and Farach-Colton, Martin and Johnson, Rob},
title = {Vector Quotient Filters: Overcoming the Time/Space Trade-Off in Filter Design},
year = {2021},
isbn = {9781450383431},
publisher = {Association for Computing Machinery},
address = {New York, NY, USA},
url = {https://doi.org/10.1145/3448016.3452841},
doi = {10.1145/3448016.3452841},
booktitle = {Proceedings of the 2021 International Conference on Management of Data},
pages = {1386–1399},
numpages = {14},
location = {Virtual Event, China},
series = {SIGMOD '21}
}

@String{BIT = "{BIT}" }

@String{Computing = "Computing" }

@String{Computer = "{IEEE} Computer" }

@String{Springer = "Springer-Verlag" }

@phdthesis{2,
title={A Highly Productive Implementation of an Out-of-Order Processor Generator},
author={Celio, Christopher P.},
school={University of California, Berkeley},
year={2017},
}

@inproceedings{GPUMEMORYWALL,
author = {Gao, Yanjie and Liu, Yu and Zhang, Hongyu and Li, Zhengxian and Zhu, Yonghao and Lin, Haoxiang and Yang, Mao},
title = {Estimating GPU memory consumption of deep learning models},
year = {2020},
isbn = {9781450370431},
publisher = {Association for Computing Machinery},
address = {New York, NY, USA},
url = {https://doi.org/10.1145/3368089.3417050},
doi = {10.1145/3368089.3417050},
booktitle = {Proceedings of the 28th ACM Joint Meeting on European Software Engineering Conference and Symposium on the Foundations of Software Engineering},
pages = {1342–1352},
numpages = {11},
location = {Virtual Event, USA},
series = {ESEC/FSE 2020}
}

@inproceedings{scheduling4,
author = {Koo, Gunjae and Oh, Yunho and Ro, Won Woo and Annavaram, Murali},
title = {Access Pattern-Aware Cache Management for Improving Data Utilization in GPU},
year = {2017},
isbn = {9781450348928},
publisher = {Association for Computing Machinery},
url = {https://doi.org/10.1145/3079856.3080239},
doi = {10.1145/3079856.3080239},
booktitle = {Proceedings of the 44th Annual International Symposium on Computer Architecture},
pages = {307–319},
numpages = {13},
series = {ISCA '17}
}

@online{A100_white_paper,
  author = {NVIDIA},
  title={NVIDIA A100 Tensor Core GPU Architecture},
  year = {2020},
  url = {https://images.nvidia.com/aem-dam/en-zz/Solutions/data-center/nvidia-ampere-architecture-whitepaper.pdf},
  urldate = {2024-06-22},
  publisher = {NVIDIA}
}

@online{ldgsts, 
  author = {NVIDIA},
  title={Parallel Thread Execution ISA Version 8.5},
  year = {2024},
  url = { https://docs.nvidia.com/cuda/parallel-thread-execution/index.html#data-movement-and-conversion-instructions-cp-async},
  urldate = {2024-06-22},
  publisher = {NVIDIA}
}

@online{H100_white_paper,
  author = {NVIDIA},
  title={NVIDIA H100 Tensor Core GPU Architecture},
  year = {2022},
  url = {https://resources.nvidia.com/en-us-tensor-core},
  urldate = {2024-06-22},
  publisher = {NVIDIA}
}

@online{GB200_white_paper,
  author = {NVIDIA},
  title={NVIDIA Blackwell Architecture Technical Brief},
  year = {2024},
  url = {https://resources.nvidia.com/en-us-blackwell-architecture},
  urldate = {2024-06-22},
  publisher = {NVIDIA}
}

@online{V100_white_paper,
  author = {NVIDIA},
  title={NVIDIA TESLA V100 GPU ARCHITECTURE},
  year = {2017},
  url = {https://images.nvidia.com/content/volta-architecture/pdf/volta-architecture-whitepaper.pdf},
  urldate = {2024-06-22},
  publisher = {NVIDIA}
}

@INPROCEEDINGS{gto,
  author={Rogers, Timothy G. and O'Connor, Mike and Aamodt, Tor M.},
  booktitle={2012 45th Annual IEEE/ACM International Symposium on Microarchitecture}, 
  title={Cache-Conscious Wavefront Scheduling}, 
  year={2012},
  volume={},
  number={},
  pages={72-83},
  doi={10.1109/MICRO.2012.16}}

@ARTICLE{osm,
  author={Darabi, Sina and Yousefzadeh-Asl-Miandoab, Ehsan and Akbarzadeh, Negar and Falahati, Hajar and Lotfi-Kamran, Pejman and Sadrosadati, Mohammad and Sarbazi-Azad, Hamid},
  journal={IEEE Transactions on Parallel and Distributed Systems}, 
  title={OSM: Off-Chip Shared Memory for GPUs}, 
  year={2022},
  volume={33},
  number={12},
  pages={3415-3429},
  doi={10.1109/TPDS.2022.3154315}}

@inproceedings{rodinia,
author = {Che, Shuai and Boyer, Michael and Meng, Jiayuan and Tarjan, David and Sheaffer, Jeremy W. and Lee, Sang-Ha and Skadron, Kevin},
title = {Rodinia: A benchmark suite for heterogeneous computing},
year = {2009},
isbn = {9781424451562},
publisher = {IEEE Computer Society},
address = {USA},
url = {https://doi.org/10.1109/IISWC.2009.5306797},
doi = {10.1109/IISWC.2009.5306797},
booktitle = {Proceedings of the 2009 IEEE International Symposium on Workload Characterization (IISWC)},
pages = {44–54},
numpages = {11},
series = {IISWC '09}
}

@INPROCEEDINGS{accel-sim,
  author={Khairy, Mahmoud and Shen, Zhesheng and Aamodt, Tor M. and Rogers, Timothy G.},
  booktitle={2020 ACM/IEEE 47th Annual International Symposium on Computer Architecture (ISCA)}, 
  title={Accel-Sim: An Extensible Simulation Framework for Validated GPU Modeling}, 
  year={2020},
  volume={},
  number={},
  pages={473-486},
  doi={10.1109/ISCA45697.2020.00047}}

@inproceedings{accel-watch,
author = {Kandiah, Vijay and Peverelle, Scott and Khairy, Mahmoud and Pan, Junrui and Manjunath, Amogh and Rogers, Timothy G. and Aamodt, Tor M. and Hardavellas, Nikos},
title = {AccelWattch: A Power Modeling Framework for Modern GPUs},
year = {2021},
isbn = {9781450385572},
publisher = {Association for Computing Machinery},
address = {New York, NY, USA},
url = {https://doi.org/10.1145/3466752.3480063},
doi = {10.1145/3466752.3480063},
booktitle = {MICRO-54: 54th Annual IEEE/ACM International Symposium on Microarchitecture},
pages = {738–753},
numpages = {16},
location = {Virtual Event, Greece},
series = {MICRO '21}
}

@inproceedings{gpuwattch,
author = {Leng, Jingwen and Hetherington, Tayler and ElTantawy, Ahmed and Gilani, Syed and Kim, Nam Sung and Aamodt, Tor M. and Reddi, Vijay Janapa},
title = {GPUWattch: Enabling Energy Optimizations in GPGPUs},
year = {2013},
publisher = {Association for Computing Machinery},
address = {New York, NY, USA},
url = {https://doi.org/10.1145/2485922.2485964},
doi = {10.1145/2485922.2485964},
booktitle = {Proceedings of the 40th Annual International Symposium on Computer Architecture},
pages = {487--498},
numpages = {12},
series = {ISCA '13}
}

@inproceedings{booksim_noc,
  author={Nan Jiang and Becker, Daniel U. and Michelogiannakis, George and Balfour, James and Towles, Brian and Shaw, D. E. and Kim, John and Dally, William J.},
  booktitle={2013 IEEE International Symposium on Performance Analysis of Systems and Software (ISPASS)}, 
  title={A detailed and flexible cycle-accurate Network-on-Chip simulator}, 
  year={2013},
  volume={},
  number={},
  pages={86-96},
  doi={10.1109/ISPASS.2013.6557149}}

@ARTICLE{2bitpred,
  author={Nair, R.},
  journal={IEEE Transactions on Computers}, 
  title={Optimal 2-bit branch predictors}, 
  year={1995},
  volume={44},
  number={5},
  pages={698-702},
  doi={10.1109/12.381956}}

@INPROCEEDINGS{l1cacheredundancy1,
  author={Tabbakh, Abdulaziz and Annavaram, Murali and Qian, Xuehai},
  booktitle={2017 IEEE International Parallel and Distributed Processing Symposium (IPDPS)}, 
  title={Power Efficient Sharing-Aware GPU Data Management}, 
  year={2017},
  volume={},
  number={},
  pages={698-707},
  doi={10.1109/IPDPS.2017.106}}

@inproceedings{l1cacheredundancy2,
author = {Ibrahim, Mohamed Assem and Kayiran, Onur and Eckert, Yasuko and Loh, Gabriel H. and Jog, Adwait},
title = {Analyzing and Leveraging Shared L1 Caches in GPUs},
year = {2020},
isbn = {9781450380751},
publisher = {Association for Computing Machinery},
address = {New York, NY, USA},
url = {https://doi.org/10.1145/3410463.3414623},
doi = {10.1145/3410463.3414623},
booktitle = {Proceedings of the ACM International Conference on Parallel Architectures and Compilation Techniques},
pages = {161–173},
numpages = {13},
location = {Virtual Event, GA, USA},
series = {PACT '20}
}

@inproceedings{sharedmemorysoftware1,
author = {Chen, Linchuan and Agrawal, Gagan},
title = {Optimizing MapReduce for GPUs with effective shared memory usage},
year = {2012},
isbn = {9781450308052},
publisher = {Association for Computing Machinery},
address = {New York, NY, USA},
url = {https://doi.org/10.1145/2287076.2287109},
doi = {10.1145/2287076.2287109},
booktitle = {Proceedings of the 21st International Symposium on High-Performance Parallel and Distributed Computing},
pages = {199–210},
numpages = {12},
location = {Delft, The Netherlands},
series = {HPDC '12}
}

@INPROCEEDINGS{sharedmemorysoftware2,
  author={Moazeni, Maryam and Bui, Alex and Sarrafzadeh, Majid},
  booktitle={2009 IEEE 7th Symposium on Application Specific Processors}, 
  title={A memory optimization technique for software-managed scratchpad memory in GPUs}, 
  year={2009},
  volume={},
  number={},
  pages={43-49},
  doi={10.1109/SASP.2009.5226334}}

@inproceedings{stash,
author = {Komuravelli, Rakesh and Sinclair, Matthew D. and Alsop, Johnathan and Huzaifa, Muhammad and Kotsifakou, Maria and Srivastava, Prakalp and Adve, Sarita V. and Adve, Vikram S.},
title = {Stash: have your scratchpad and cache it too},
year = {2015},
isbn = {9781450334020},
publisher = {Association for Computing Machinery},
address = {New York, NY, USA},
url = {https://doi.org/10.1145/2749469.2750374},
doi = {10.1145/2749469.2750374},
booktitle = {Proceedings of the 42nd Annual International Symposium on Computer Architecture},
pages = {707–719},
numpages = {13},
location = {Portland, Oregon},
series = {ISCA '15}
}

@article{cacti,
  title={CACTI 6.0: A tool to model large caches},
  author={Muralimanohar, Naveen and Balasubramonian, Rajeev and Jouppi, Norman P and others},
  journal={HP laboratories},
  volume={27},
  pages={28},
  year={2009},
  url = {https://shiftleft.com/mirrors/www.hpl.hp.com/techreports/2009/HPL-2009-85.pdf}
  
}

@ARTICLE{sharingl1-ATA,
  author={Xu, Xiangrong and Wang, Liang and Xiao, Limin and Liu, Lei and Lv, Yuanqiu and Xie, Xilong and Han, Meng and Liu, Hao},
  journal={IEEE Transactions on Computer-Aided Design of Integrated Circuits and Systems}, 
  title={ATA-Cache: Contention Mitigation for GPU Shared L1 Cache With Aggregated Tag Array}, 
  year={2024},
  volume={43},
  number={5},
  pages={1429-1441},
  doi={10.1109/TCAD.2023.3337192}}

@article{sharingl1-CCD,
author = {Falahati, Hajar and Sadrosadati, Mohammad and Xu, Qiumin and G\'{o}mez-Luna, Juan and Saber Latibari, Banafsheh and Jeon, Hyeran and Hesaabi, Shaahin and Sarbazi-Azad, Hamid and Mutlu, Onur and Annavaram, Murali and Pedram, Masoud},
title = {Cross-core Data Sharing for Energy-efficient GPUs},
year = {2024},
issue_date = {September 2024},
publisher = {Association for Computing Machinery},
address = {New York, NY, USA},
volume = {21},
number = {3},
issn = {1544-3566},
url = {https://doi.org/10.1145/3653019},
doi = {10.1145/3653019},
journal = {ACM Trans. Archit. Code Optim.},
month = sep,
articleno = {42},
numpages = {32}
}

@article{sharingl1-RING,
author = {Dublish, Saumay and Nagarajan, Vijay and Topham, Nigel},
title = {Cooperative Caching for GPUs},
year = {2016},
issue_date = {December 2016},
publisher = {Association for Computing Machinery},
address = {New York, NY, USA},
volume = {13},
number = {4},
issn = {1544-3566},
url = {https://doi.org/10.1145/3001589},
doi = {10.1145/3001589},
journal = {ACM Trans. Archit. Code Optim.},
month = dec,
articleno = {39},
numpages = {25}
}

@INPROCEEDINGS{sharingl1-MeshL1sharing,
  author={Ibrahim, Mohamed Assem and Liu, Hongyuan and Kayiran, Onur and Jog, Adwait},
  booktitle={2019 28th International Conference on Parallel Architectures and Compilation Techniques (PACT)}, 
  title={Analyzing and Leveraging Remote-Core Bandwidth for Enhanced Performance in GPUs}, 
  year={2019},
  volume={},
  number={},
  pages={258-271},
  doi={10.1109/PACT.2019.00028}}

@inproceedings{sharingl1-L1.5Dcache,
author = {Wang, Jianfei and Jiang, Li and Ke, Jing and Liang, Xiaoyao and Jing, Naifeng},
title = {A sharing-aware L1.5D cache for data reuse in GPGPUs},
year = {2019},
isbn = {9781450360074},
publisher = {Association for Computing Machinery},
address = {New York, NY, USA},
url = {https://doi.org/10.1145/3287624.3287633},
doi = {10.1145/3287624.3287633},
booktitle = {Proceedings of the 24th Asia and South Pacific Design Automation Conference},
pages = {388–393},
numpages = {6},
location = {Tokyo, Japan},
series = {ASPDAC '19}
}

@INPROCEEDINGS{sharingl1-decouplel1,
  author={Ibrahim, Mohamed Assem and Kayiran, Onur and Eckert, Yasuko and Loh, Gabriel H. and Jog, Adwait},
  booktitle={2021 IEEE International Symposium on High-Performance Computer Architecture (HPCA)}, 
  title={Analyzing and Leveraging Decoupled L1 Caches in GPUs}, 
  year={2021},
  volume={},
  number={},
  pages={467-478},
  doi={10.1109/HPCA51647.2021.00047}}

@inproceedings{sharingl1-colab,
author = {Cheng, Bo-Wun and Huang, En-Ming and Chao, Chen-Hao and Sun, Wei-Fang and Yeh, Tsung-Tai and Lee, Chun-Yi},
title = {COLAB: Collaborative and Efficient Processing of Replicated Cache Requests in GPU},
year = {2023},
isbn = {9781450397834},
publisher = {Association for Computing Machinery},
address = {New York, NY, USA},
url = {https://doi.org/10.1145/3566097.3567838},
doi = {10.1145/3566097.3567838},
booktitle = {Proceedings of the 28th Asia and South Pacific Design Automation Conference},
pages = {314–319},
numpages = {6},
location = {Tokyo, Japan},
series = {ASPDAC '23}
}

@ARTICLE{sharingl1-IntergroupCacheCooperation,
  author={Wang, Guosheng and Du, Yajuan and Huang, Weiming},
  journal={IEEE Transactions on Computer-Aided Design of Integrated Circuits and Systems}, 
  title={GPU Performance Optimization via Intergroup Cache Cooperation}, 
  year={2024},
  volume={43},
  number={11},
  pages={4142-4153},
  doi={10.1109/TCAD.2024.3443707}}

@INPROCEEDINGS{sharingl1-Pseudo-Cache,
  author={Li, Bingchao and Wei, Jizeng and Zhu, Yuchen},
  booktitle={2024 IEEE International Conference on High Performance Computing and Communications (HPCC)}, 
  title={Pseudo-Cache: Extending the Access Scope of Requests with Global Perspective in GPUs}, 
  year={2024},
  volume={},
  number={},
  pages={50-60},
  doi={10.1109/HPCC64274.2024.00018}}

@INPROCEEDINGS{sharingl1-Collaborative-Coalescing,
  author={Jiang, Fan and Li, Chengeng and Zhang, Wei and Xu, Jiang},
  booktitle={2024 29th Asia and South Pacific Design Automation Conference (ASP-DAC)}, 
  title={Collaborative Coalescing of Redundant Memory Access for GPU System}, 
  year={2024},
  volume={},
  number={},
  pages={195-200},
  doi={10.1109/ASP-DAC58780.2024.10473837}}

@article{sharingl1-two-levelshareing,
title = {Exploiting intra-chip locality for multi-chip GPUs via two-level shared L1 cache},
journal = {Journal of Systems Architecture},
volume = {167},
pages = {103500},
year = {2025},
issn = {1383-7621},
doi = {https://doi.org/10.1016/j.sysarc.2025.103500},
url = {https://www.sciencedirect.com/science/article/pii/S1383762125001729},
author = {Xiangrong Xu and Liang Wang and Limin Xiao and Lei Liu and Zihao Zhou and Yuanqiu Lv and Li Ruan and Xilong Xie and Meng Han and Xiaojian Liao}
}

@article{BFlatencyFPGA,
title = {Hardware-oriented optimization of Bloom filter algorithms and architectures for ultra-high-speed lookups in network applications},
journal = {Microprocessors and Microsystems},
volume = {93},
pages = {104619},
year = {2022},
issn = {0141-9331},
doi = {https://doi.org/10.1016/j.micpro.2022.104619},
url = {https://www.sciencedirect.com/science/article/pii/S0141933122001582},
author = {Arish Sateesan and Jo Vliegen and Joan Daemen and Nele Mentens}
}

@article{BloomFilter,
author = {Bloom, Burton H.},
title = {Space/time trade-offs in hash coding with allowable errors},
year = {1970},
issue_date = {July 1970},
publisher = {Association for Computing Machinery},
address = {New York, NY, USA},
volume = {13},
number = {7},
issn = {0001-0782},
url = {https://doi.org/10.1145/362686.362692},
doi = {10.1145/362686.362692},
journal = {Commun. ACM},
month = jul,
pages = {422–426},
numpages = {5}
}

@INPROCEEDINGS{Morpheus,
  author={Darabi, Sina and Sadrosadati, Mohammad and Akbarzadeh, Negar and Lindegger, Joël and Hosseini, Mohammad and Park, Jisung and Gómez-Luna, Juan and Mutlu, Onur and Sarbazi-Azad, Hamid},
  booktitle={2022 55th IEEE/ACM International Symposium on Microarchitecture (MICRO)}, 
  title={Morpheus: Extending the Last Level Cache Capacity in GPU Systems Using Idle GPU Core Resources}, 
  year={2022},
  volume={},
  number={},
  pages={228-244},
  doi={10.1109/MICRO56248.2022.00029}}

@INPROCEEDINGS{BF-arch1-cache-partitioning,
  author={Nikas, Konstantinos and Horsnell, Matthew and Garside, Jim},
  booktitle={2008 International Conference on Embedded Computer Systems: Architectures, Modeling, and Simulation}, 
  title={An adaptive bloom filter cache partitioning scheme for multicore architectures}, 
  year={2008},
  volume={},
  number={},
  pages={25-32},
  doi={10.1109/ICSAMOS.2008.4664843}}

@inproceedings{BF-arch-coherence,
author = {Zebchuk, Jason and Srinivasan, Vijayalakshmi and Qureshi, Moinuddin K. and Moshovos, Andreas},
title = {A tagless coherence directory},
year = {2009},
isbn = {9781605587981},
publisher = {Association for Computing Machinery},
address = {New York, NY, USA},
url = {https://doi.org/10.1145/1669112.1669166},
doi = {10.1145/1669112.1669166},
booktitle = {Proceedings of the 42nd Annual IEEE/ACM International Symposium on Microarchitecture},
pages = {423–434},
numpages = {12},
location = {New York, New York},
series = {MICRO 42}
}

@manual{VCU118,
  author = {{AMD}},
  title = {{VCU118 Evaluation Board User Guide}},
  organization = {{AMD}},
  year = {2023},
  note = {{UG1224}},
  url = {https://docs.amd.com/v/u/en-US/ug1224-vcu118-eval-bd}
}

@article{stratton2012parboil,
  title={Parboil: A revised benchmark suite for scientific and commercial throughput computing},
  author={Stratton, John A and Rodrigues, Christopher and Sung, I-Jui and Obeid, Nady and Chang, Li-Wen and Anssari, Nasser and Liu, Geng Daniel and Hwu, Wen-mei W},
  journal={Center for Reliable and High-Performance Computing},
  volume={127},
  number={7.2},
  pages={27},
  year={2012},
  url = {http://impact.crhc.illinois.edu/Shared/Report/impact-12-01.parboil.pdf}
}

@INPROCEEDINGS{ISPASS,
  author={Bakhoda, Ali and Yuan, George L. and Fung, Wilson W. L. and Wong, Henry and Aamodt, Tor M.},
  booktitle={2009 IEEE International Symposium on Performance Analysis of Systems and Software}, 
  title={Analyzing CUDA workloads using a detailed GPU simulator}, 
  year={2009},
  volume={},
  number={},
  pages={163-174},
  doi={10.1109/ISPASS.2009.4919648}}

@INPROCEEDINGS{pannotia,
  author={Che, Shuai and Beckmann, Bradford M. and Reinhardt, Steven K. and Skadron, Kevin},
  booktitle={2013 IEEE International Symposium on Workload Characterization (IISWC)}, 
  title={Pannotia: Understanding irregular GPGPU graph applications}, 
  year={2013},
  volume={},
  number={},
  pages={185-195},
  doi={10.1109/IISWC.2013.6704684}}

@INPROCEEDINGS{polybenchGPU,
  author={Grauer-Gray, Scott and Xu, Lifan and Searles, Robert and Ayalasomayajula, Sudhee and Cavazos, John},
  booktitle={2012 Innovative Parallel Computing (InPar)}, 
  title={Auto-tuning a high-level language targeted to GPU codes}, 
  year={2012},
  volume={},
  number={},
  pages={1-10},
  doi={10.1109/InPar.2012.6339595}}

@inproceedings {flush_reload,
author = {Yuval Yarom and Katrina Falkner},
title = {{FLUSH+RELOAD}: A High Resolution, Low Noise, L3 Cache {Side-Channel} Attack},
booktitle = {23rd USENIX Security Symposium (USENIX Security 14)},
year = {2014},
isbn = {978-1-931971-15-7},
address = {San Diego, CA},
pages = {719--732},
url = {https://www.usenix.org/conference/usenixsecurity14/technical-sessions/presentation/yarom},
publisher = {USENIX Association},
month = aug
}

@InProceedings{gpu_cache_attack,
author="Moghimi, Ahmad
and Irazoqui, Gorka
and Eisenbarth, Thomas",
editor="Fischer, Wieland
and Homma, Naofumi",
title="CacheZoom: How SGX Amplifies the Power of Cache Attacks",
booktitle="Cryptographic Hardware and Embedded Systems -- CHES 2017",
year="2017",
publisher="Springer International Publishing",
address="Cham",
pages="69--90",
isbn="978-3-319-66787-4"
}

@article{bloom_finger_Sal,
title = {Improving counting Bloom filter performance with fingerprints},
journal = {Information Processing Letters},
volume = {116},
number = {4},
pages = {304-309},
year = {2016},
issn = {0020-0190},
doi = {https://doi.org/10.1016/j.ipl.2015.11.002},
url = {https://www.sciencedirect.com/science/article/pii/S0020019015001866},
author = {Salvatore Pontarelli and Pedro Reviriego and Juan Antonio Maestro}
}

@inproceedings{BFFPGA2,
author = {Seshadri, Vivek and Mutlu, Onur and Kozuch, Michael A. and Mowry, Todd C.},
title = {The evicted-address filter: a unified mechanism to address both cache pollution and thrashing},
year = {2012},
isbn = {9781450311823},
publisher = {Association for Computing Machinery},
address = {New York, NY, USA},
url = {https://doi.org/10.1145/2370816.2370868},
doi = {10.1145/2370816.2370868},
booktitle = {Proceedings of the 21st International Conference on Parallel Architectures and Compilation Techniques},
pages = {355–366},
numpages = {12},
location = {Minneapolis, Minnesota, USA},
series = {PACT '12}
}

@inproceedings{bfNUCA,
  title={Leveraging bloom filters for smart search within NUCA caches},
  author={Ricci, Robert and Barrus, Steve and Gebhardt, Dan and Balasubramonian, Rajeev},
  booktitle={7th Workshop on Complexity-Effective Design (WCED)},
  year={2006},
  url = {https://doi.org/10.1145/2370816.2370868},
  doi={https://www-old.cs.utah.edu/~rajeev/pubs/wced06.pdf}
}

@INPROCEEDINGS{BFarchL2tag,
  author={Park, Hyunsun and Yoo, Sungjoo and Lee, Sunggu},
  booktitle={2011 Design, Automation \& Test in Europe}, 
  title={A novel tag access scheme for low power L2 cache}, 
  year={2011},
  volume={},
  number={},
  pages={1-6},
  doi={10.1109/DATE.2011.5763108}}

@INPROCEEDINGS{BFarchcachemiss,
  author={Tao, Xi and Zeng, Qi and Peir, Jih-Kwon and Lu, Shih-Lien},
  booktitle={2016 17th International Conference on Parallel and Distributed Computing, Applications and Technologies (PDCAT)}, 
  title={Runahead Cache Misses Using Bloom Filter}, 
  year={2016},
  volume={},
  number={},
  pages={1-6},
  doi={10.1109/PDCAT.2016.017}}

@inproceedings{bfarchcacheprefetching,
author = {Peir, Jih-Kwon and Lai, Shih-Chang and Lu, Shih-Lien and Stark, Jared and Lai, Konrad},
title = {Bloom filtering cache misses for accurate data speculation and prefetching},
year = {2002},
isbn = {9781450328401},
publisher = {Association for Computing Machinery},
address = {New York, NY, USA},
url = {https://doi.org/10.1145/2591635.2667183},
doi = {10.1145/2591635.2667183},
booktitle = {ACM International Conference on Supercomputing 25th Anniversary Volume},
pages = {347–356},
numpages = {10},
location = {Munich, Germany}
}

@article{bffpga,
author = {Putze, Felix and Sanders, Peter and Singler, Johannes},
title = {Cache-, hash-, and space-efficient bloom filters},
year = {2010},
issue_date = {2009},
publisher = {Association for Computing Machinery},
address = {New York, NY, USA},
volume = {14},
issn = {1084-6654},
url = {https://doi.org/10.1145/1498698.1594230},
doi = {10.1145/1498698.1594230},
journal = {ACM J. Exp. Algorithmics},
month = jan,
articleno = {4},
numpages = {18}
}

@INPROCEEDINGS{bfarchcachepollution,
  author={Ashihara, Takakazu and Kamiyama, Noriaki},
  booktitle={2021 IEEE International Symposium on Local and Metropolitan Area Networks (LANMAN)}, 
  title={Detecting Cache Pollution Attacks Using Bloom Filter}, 
  year={2021},
  volume={},
  number={},
  pages={1-6},
  doi={10.1109/LANMAN52105.2021.9478804}}

@inproceedings{bfarchvirtual_cache_synonym_lookup,
author = {Woo, Dong Hyuk and Ghosh, Mrinmoy and \"{O}zer, Emre and Biles, Stuart and Lee, Hsien-Hsin S.},
title = {Reducing energy of virtual cache synonym lookup using bloom filters},
year = {2006},
isbn = {1595935436},
publisher = {Association for Computing Machinery},
address = {New York, NY, USA},
url = {https://doi.org/10.1145/1176760.1176783},
doi = {10.1145/1176760.1176783},
booktitle = {Proceedings of the 2006 International Conference on Compilers, Architecture and Synthesis for Embedded Systems},
pages = {179–189},
numpages = {11},
location = {Seoul, Korea},
series = {CASES '06}
}

@article{bfstructurecache,
  title={Cache efficient bloom filters for shared memory machines},
  author={Kaler, Tim},
  year = {2013},
  
  url = {http://tfk.mit.edu/pdf/bloom.pdf}
}

@inproceedings{bfcachesharing_network,
author = {Mun, Ju Hyoung and Lim, Hyesook},
title = {Cache Sharing Using a Bloom Filter in Named Data Networking},
year = {2016},
isbn = {9781450341837},
publisher = {Association for Computing Machinery},
address = {New York, NY, USA},
url = {https://doi.org/10.1145/2881025.2889477},
doi = {10.1145/2881025.2889477},
booktitle = {Proceedings of the 2016 Symposium on Architectures for Networking and Communications Systems},
pages = {127–128},
numpages = {2},
location = {Santa Clara, California, USA},
series = {ANCS '16}
}

@article{bloomfilterFP,
author = {Andrei Broder and Michael Mitzenmacher},
title = {Network Applications of Bloom Filters: A Survey},
journal = {Internet Mathematics},
volume = {1},
number = {4},
pages = {485--509},
year = {2004},
publisher = {Taylor \& Francis},
doi = {10.1080/15427951.2004.10129096},
URL = {https://doi.org/10.1080/15427951.2004.10129096},
eprint = { https://doi.org/10.1080/15427951.2004.10129096}

}

@misc{kvcache2026,
  author = {Ganjihal, Sanjeev Rao},
  title = {Predictive Multi-Tier Memory Management for KV Cache in Large-Scale GPU Inference},
  year = {2026},
  eprint = {2604.26968},
  archivePrefix = {arXiv},
  primaryClass = {cs.DC},
  url = {https://arxiv.org/abs/2604.26968}
}

@misc{neo2024,
  author = {Choi, Wonbeom and Baek, Jungi and Park, Seunghyun and Yoon, Jeongmin and Park, Jaehong and Kwon, Ikhyeon and Bae, Jonghyun and Lee, Jangwoo},
  title = {{NEO}: Saving GPU Memory Crisis with CPU Offloading for Online LLM Inference},
  year = {2024},
  eprint = {2411.01142},
  archivePrefix = {arXiv},
  primaryClass = {cs.DC},
  url = {https://arxiv.org/abs/2411.01142}
}

@misc{wang2020crosslayer,
  author = {Wang, Yawen and Gong, Chengyue and Gao, Zhe and Li, Ang and Jiang, Xuehai and Chen, Xiang},
  title = {Accelerating Deep Learning Inference with Cross-Layer Data Reuse on GPUs},
  year = {2020},
  eprint = {2007.06000},
  archivePrefix = {arXiv},
  primaryClass = {cs.LG},
  url = {https://arxiv.org/abs/2007.06000}
}

\end{document}